\documentclass[aps,prapplied,reprint,superscriptaddress,showpacs,amsmath]{revtex4-2}
\usepackage{lmodern}
\usepackage{lipsum}
\usepackage{physics}

\usepackage{lmodern}
\usepackage[T1]{fontenc}
\usepackage{lipsum, color}
\usepackage{amstext}
\usepackage{booktabs}
\usepackage{tabularx} 
\usepackage{amssymb}
\usepackage{makecell}
\usepackage{amsmath}
\usepackage{xfrac}
\usepackage{esint}
\usepackage{array}
\usepackage{tablefootnote} 
\newcommand{\proptosim}{\mathrel{\vcenter{\offinterlineskip
\halign{\hfil$##$\cr
\propto\cr
\sim\cr}}}}

\newcommand{\etal}{\textit{et al.}}
\usepackage{graphicx}
\newcommand{\appropto}{\mathrel{\vcenter{
  \offinterlineskip\halign{\hfil$##$\cr
  \propto\cr\noalign{\kern0pt}\sim\cr}}}}
\usepackage{setspace}   

\begin{document}
\title{Broadband Kinetic-Inductance Parametric Amplifiers with Impedance Engineering }

\author{Chih-Chiao Hung} 
\email{chih-chiao.hung@riken.jp}
\affiliation{RIKEN Center for Quantum Computing (RQC), Wako, Saitama 351-0198, Japan}
\author{Hiroki Kutsuma} 
\affiliation{Graduate School of Engineering, Tohoku University, Sendai 980-8579, Japan}
\author{Chung~Wai~Sandbo~Chang} 
\affiliation{RIKEN Center for Quantum Computing (RQC), Wako, Saitama 351-0198, Japan}
\author{Arjan~Ferdinand~van~Loo} 
\affiliation{RIKEN Center for Quantum Computing (RQC), Wako, Saitama 351-0198, Japan}
\affiliation{Department of Applied Physics, Graduate School of Engineering, The University of Tokyo, Bunkyo-ku, Tokyo 113-8656, Japan}
\author{Yasunobu Nakamura} 
\email{yasunobu@ap.t.u-tokyo.ac.jp}
\affiliation{RIKEN Center for Quantum Computing (RQC), Wako, Saitama 351-0198, Japan}
\affiliation{Department of Applied Physics, Graduate School of Engineering, The University of Tokyo, Bunkyo-ku, Tokyo 113-8656, Japan}

\date{\today}
\begin{abstract}
    Broadband quantum-limited parametric amplifiers~(PAs) are essential components in quantum information science and technology. 
    Impedance-engineered resonator-based PAs and traveling-wave PAs are the primary approaches to overcome the gain--bandwidth constraint. 
    While the former PAs are simpler to fabricate, the target characteristic impedance $Z_\mathrm{NR}$ of the nonlinear resonator has been restricted to be below 10~$\Omega$, requiring large capacitance.
    Moreover, these PAs have only been implemented with aluminum-based Josephson junctions~(JJs), hindering their operation at high temperatures or strong magnetic fields.
    To address these issues, we propose a three-stage impedance-transformer scheme, showcased with a 20-nm-thick, 250-nm-wide high-kinetic-inductance niobium-titanium-nitride~(NbTiN) film.
    Our scheme enables $Z_\mathrm{NR}$ up to several tens of ohms—a tenfold improvement over conventional designs, achieved through an additional quarter-wavelength transmission line with the characteristic impedance of 180~$\Omega$.
    Our kinetic-inductance impedance-engineered parametric amplifiers~(KIMPA), featuring a 330-fF shunt capacitor, demonstrate a phase-preserving amplification with a 450-MHz bandwidth at 17-dB gain, and an added noise ranging from 0.5--1.3 quanta near the center frequency of 8.4~GHz. 
    Due to the high critical current of the NbTiN nanowire, the KIMPA also achieves an output saturation power of up to $-$51$\pm$3~dBm, approximately 25-dB higher than that of JJ-based PAs. 
    This scheme also opens new possibilities for other three-wave-mixing building blocks.
\end{abstract}
\maketitle

\section{Introduction}

Quantum-noise-limited microwave parametric amplifiers~(PAs), which exploit parametric photon conversion, have become indispensable for sensitive quantum device readouts in multiple scientific fields~\cite{vijay2009invited,clerk2010introduction,macklin2015near,aumentado2020superconducting}.
Their effectiveness stems from the strategic placement at the first signal-amplification stage, typically located at the lowest temperature, while providing sufficient gain to overcome noise from subsequent stages including high-electron-mobility-transistor~(HEMT) amplifiers.
These PAs have significantly enhanced qubit-state discrimination in superconducting quantum information research~\cite{macklin2015near,heinsoo2018rapid} and enabled the detection of previously indiscernible weak signals in high-sensitivity spin resonance experiments~\cite{bienfait2016reaching,eichler2017electron,vine2023situ}, itinerant single microwave photons~\cite{eichler2011experimental,mallet2011quantum}, and dark matter searches of axions~\cite{dark_matter_1,dark_matter_2,ahn2024extensive}.
With the progression of quantum computing toward fault-tolerant systems that necessitate concurrent multi-qubit readout~\cite{heinsoo2018rapid,spring2025fast} and error correction~\cite{corcoles2015demonstration,zhao2022realization}, there is an increasing demand for next-generation PAs with attributes of low added noise, wide bandwidth, and high saturation power.

Al-AlO$_x$-Al Josephson junctions~(JJs) have long served as amplifying elements~\cite{yurke1989observation}, offering controllable nonlinearity through oxidation conditions.
Although JJ-based PAs demonstrate near-ideal quantum efficiency~\cite{castellanos2007widely,yamamoto2008flux,roy2015broadband}, they are limited to operation below the critical temperature $T_\mathrm{c}$ of aluminum and in low magnetic fields.
Moreover, their input 1-dB compression power $P_{1\mathrm{dB}}^{\mathrm{in}}$ ranges from $-100$ to $-93$ dBm or the output saturation power of approximately $-75$ dBm~\cite{macklin2015near,ezenkova2022broadband,kaufman2023josephson} due to relatively low critical currents and higher-order nonlinearities~\cite{boutin2017effect}, presenting challenges for simultaneous readouts of tens of qubits~\cite{remm2023intermodulation}.

In contrast, kinetic-inductance~(KI) materials such as NbN and NbTiN offer distinct advantages: high operating temperature~\cite{ho2012wideband,malnou2022performance,frasca2024three} due to high $T_\mathrm{c}$~\cite{zmuidzinas2012superconducting} and high $P_{1\mathrm{dB}}^{\mathrm{in}}$~($-$65 to $-$55~dBm)~\cite{ho2012wideband,bockstiegel2014development,shu2021nonlinearity,parker2022degenerate} attributed to larger critical currents compared to JJs~\cite{vissers2015frequency}.
Additionally, their larger superconducting gaps~\cite{makise2010characterization,niepce2019high,frasca2024three} enable better performance in strong magnetic fields~\cite{samkharadze2016high,vine2023situ} and potentially higher operating temperatures and frequencies. 
Generally, the lower intrinsic nonlinearity of KI materials~\cite{parker2022degenerate} necessitates higher pump powers to achieve comparable bandwidth and gain to JJ-based designs.
As such, JJ-based PAs might be better for some applications, while the choice between JJ- and KI-based PAs involves operating conditions, required dynamic ranges, and pump-power constraints, making these trade-offs crucial for amplifier design.

The pioneering PA generations exhibited a fixed gain--bandwidth product~(GBWP)~\cite{castellanos2007widely,yamamoto2008flux}, where increased gain came at the cost of reduced operational frequency range~(bandwidth).
Recent efforts to overcome this limitation have followed two main approaches. 
Traveling-wave parametric amplifiers~(TWPAs), utilizing thousands of nonlinear unit cells as transmission lines, have achieved bandwidths approaching 4~GHz~\cite{bockstiegel2014development,macklin2015near}. 
However, they suffer from gain ripples due to the impedance mismatches~\cite{bockstiegel2014development,faramarzi20244} and exhibit relatively low quantum efficiency compared to conventional PAs due to the higher dissipation of lossy dielectrics~\cite{peng2022floquet}.
An alternative approach utilizes passive networks like Klopfenstein tapers~\cite{mutus2014strong,lu2022broadband,white2023readout,qing2024broadband} or impedance transformers~\cite{roy2015broadband,grebel2021flux,ezenkova2022broadband,ranzani2022wideband,naaman2022synthesis} to reduce the external quality factor of lumped-element nonlinear oscillators, which operate in the reflection mode and require circulators. 
These designs, demanding minimal fabrication resources, attain flat-top gain profiles with small ripples and near-quantum-limited noise performance.
However, these impedance-engineered PAs typically require large shunt capacitances of several picofarads and sometimes employ parallel-plate capacitors with possible additional loss from the dielectrics~\cite{mutus2014strong,roy2015broadband,grebel2021flux}~(see Appendix~\ref{Appedix_roy}).
Furthermore, they are limited to JJs rather than KI materials.
We compare different reflection-type JJ-based broadband PAs along with narrowband KI PAs in Appendix~\ref{appen_table}. 

\begin{figure}
\centering{}
\includegraphics[width=8.27cm]{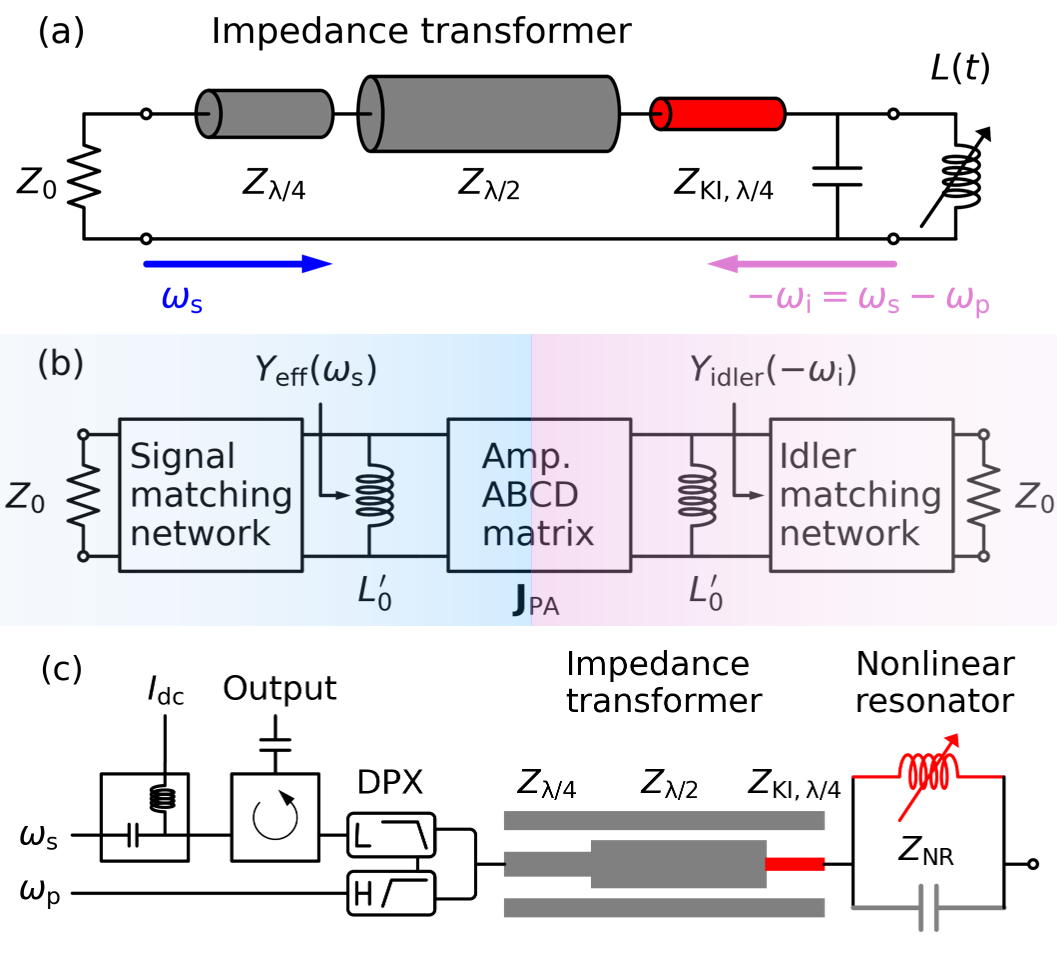}
    \caption{
    (a) Circuit diagram of our kinetic-inductance impedance-engineered parametric amplifier (KIMPA).
    The three-stage impedance transformer, consisting of two quarter-wavelength~($\lambda/4$) and one half-wavelength~($\lambda/2$) transmission lines, broadens the bandwidth of the nonlinear resonator~(NR) consisting of a shunt capacitor and a modulated inductor $L(t)$.
    The resonance of NR is close to the resonance of the $\lambda/2$ line.
    (b)~Equivalent time-independent circuit diagram.
    The amplification ABCD matrix describes the relationship of signal and idler currents, sent from the left and the right, respectively.
    The signal~(blue) and the idler~(lavender) segments undergo frequencies of $\omega_\mathrm{s}$ and $-\omega_\mathrm{i}$, respectively.
    (c) Schematic diagram of~(a) on the right and the experimental setup on the left. 
    The signal path incorporates a bias tee for the dc-current~($I_\mathrm{dc}$) injection and a diplexer for combining the signal and pump. 
    The device is fabricated using two materials, NbTiN~(red) and aluminum~(gray).
    The nonlinear resonator has a characteristic impedance $Z_{\mathrm{NR}}$.}
    \label{fig0}
\end{figure}

Here, we propose an impedance-engineering scheme using a three-stage impedance transformer~[Fig.~\ref{fig0}(a)] to reduce the shunt capacitance to a few-hundred femtofarads while achieving a wide bandwidth and maintaining a simple fabrication recipe.
We demonstrate this approach using a NbTiN thin film and discuss its nonlinear properties in Sec.~\ref{sec.NbTiN}, along with fabrication details in Sec.~\ref{sec_fab}.
Performance of the KI-based impedance-engineered PA~(KIMPA) is presented in Sec.~\ref{sec_measurement}: 17-dB signal gain, 450-MHz bandwidth, $-$51$\pm$3-dBm output saturation power, and 0.5--1.3-quanta added noise.  

\section{Three-stage Impedance transformer}
\label{sec.three}
The nonlinearity of PAs under pumped conditions can be modeled by treating the impedance of the nonlinear element~(typically an inductor) as a purely imaginary component in parallel with an effective negative resistance~\cite{sundqvist2014negative,naaman2022synthesis}, resulting in gain~\cite{getsinger1963prototypes}.
The time-dependent inductance modulated at the pump frequency $\omega_\mathrm{p}$ follows
\begin{equation}
    L(t) = L_0 + \frac{1}{2}\left(\delta L\,e^{i \omega_\mathrm{p}t}+ \mathrm{c.c.}\right), \label{eq_mod_L}
\end{equation}
where $L_0$ is the unmodulated inductance, and $\delta L$ denotes the complex modulation amplitude.
A nonlinear resonator~(NR) consisting of a time-dependent $L(t)$ in parallel with a capacitor enables three-wave-mixing~(3WM) parametric amplification by converting a pump photon at $\omega_\mathrm{p}$ into a signal photon at $\omega_\mathrm{s}$ and an idler photon at $\omega_\mathrm{i}$, when $\omega_\mathrm{p} = \omega_\mathrm{s} + \omega_\mathrm{i}$.
For the examination of non-degenerate amplification~($\omega_\mathrm{s}\neq\omega_\mathrm{i}$), we aim to replace the time-dependent $L(t)$ with an effective time-independent inductor $L_\mathrm{eff}$. 

To determine $L_\mathrm{eff}$, $L(t)$ is initially analyzed as the central set of three elements that link the signal- and idler-port currents through the signal~(in blue) and idler~(in lavender) matching networks as illustrated in Fig.~\ref{fig0}(b)~\cite{naaman2022synthesis}.
These three elements consist of one amplification inverter ABCD matrix~($\mathbf{J}_\mathrm{PA}$), which characterizes the parametric coupling of signal and idler currents, and two parallel time-independent $L_0^\prime$ components~(for derivation details, refer to Appendix~\ref{appendix_modulate_inductor}).
Here, $L_0^\prime = L_0(1 - \alpha)$ depends on the dimensionless modulation strength $\alpha = |\delta L|^2/4L_0^2$~\cite{naaman2022synthesis}.
Note that the idler component~(right side of $\mathbf{J}_\mathrm{PA}$ in lavender) arises from the idler current at $-\omega_\mathrm{i}$, distinct from the signal counterpart experiencing $\omega_\mathrm{s}$.
Moreover, the idler matching network mirrors the signal circuit, incorporating components between the port and $L(t)$ within two pairs of open dots in Fig.~\ref{fig0}(a).
The series of matching networks of the signal and the idler is highlighted by blue and lavender arrows in Fig.~\ref{fig0}(a).

In Appendix~\ref{appendix_modulate_inductor}, we summarize the complicated circuit described above into an effective time-independent admittance $Y_\mathrm{eff}$ to represent $L(t)$ as~\cite{naaman2022synthesis}:
\begin{align}
    Y_{\mathrm{eff}}(\omega_\mathrm{s}) &= \frac{1}{i\omega_\mathrm{s}L_0^\prime}\left[1+\frac{\alpha}{i\omega_\mathrm{i} L_0^\prime Y_\mathrm{idler}^\ast(\omega_\mathrm{i}) - 1}\right]\label{eq_eff_Y}\\
    &= \frac{1}{i\omega_\mathrm{s}L_\mathrm{eff}},
\end{align}
where $Y_k$~($k = \mathrm{idler},\,\mathrm{eff}$) are defined as the admittance looking from the arrow direction in Fig.~\ref{fig0}(b).
$L_\mathrm{eff}$ is a complex number that can resemble a standard inductor parallel to a resistor with a negative value under a suitable impedance-matching network~(related to $Y_\mathrm{idler}$) and a proper pump current~(see Appendix~\ref{appendix_network}).
Since $Y_\mathrm{idler}$ is the admittance looking from the inductor to the idler port, $Y_\mathrm{idler}(-\omega_\mathrm{i}) = Y_\mathrm{idler}^\ast(\omega_\mathrm{i})$.

The bandwidth $\Delta\omega$ of PAs for any negative-resistance devices can be designed and calculated~\cite{getsinger1963prototypes}. 
In a basic signal-matching network with an NR linking to a port via a coupled capacitor~\cite{castellanos2007widely,yamamoto2008flux} or a stepped-impedance band-stop filter~\cite{parker2022degenerate}, the PA faces the GBWP limitation.
They are categorized as a single-pole PA meaning a total of one resonant mode in the system~\cite{naaman2022synthesis}.
To overcome the GBWP constraint, additional $N$($\ge 1$) coupled modes are introduced to the NR, using half-wavelength $\lambda/2$ coplanar waveguides~(CPWs)~\cite{roy2015broadband} or lumped-element resonators~\cite{kaufman2023josephson}. 
This transforms the PA into a~($N$+1)-pole system with $N$+1 local gain peaks. 
The mechanism of band-broadening in a multi-pole PA is analogous to a multi-pole band-pass impedance-matching network~(e.g., Chebyshev, Butterworth, etc.)~\cite{getsinger1963prototypes,naaman2022synthesis}; coupling multiple resonators enables controlled pole placement, resulting in broader and flatter gain or filtering.

We introduce a three-stage impedance-transformer matching network as an example of a two-pole system, depicted in Fig.~\ref{fig0}(a). 
This two-pole PA features a $\lambda/2$ CPW resonator, an NR, and two $\lambda/4$ CPWs acting as an impedance inverter. 
Unlike the conventional setup in Ref.~\citenum{roy2015broadband}, which lacks the red $\lambda/4$ CPW in Fig.~\ref{fig0}(a), our design supports a broader range for the characteristic impedance $Z_{\mathrm{NR}}$ of a NR, necessitating a shunt capacitance of a few-hundred femtofarads rather than several picofarads.
The key improvement of our scheme is this additional $\lambda/4$ CPW with a high impedance of $Z_{\mathrm{KI},\lambda/4} = 180\ \Omega$, placed between the $\lambda/2$ CPW resonator and the NR. 
Achieving such a high-impedance CPW involves specialized methods, like using the kinetic inductance of NbTiN, since $Z_{\mathrm{KI},\lambda/4}$ exceeds the standard CPW geometric-impedance limit of about 90~$\Omega$~(for a 1\nobreakdash-$\mu$m-wide center line and 20-$\mu$m gap).
In our design, we set $Z_{\mathrm{NR}} = 56\ \Omega$~(corresponding to a 330-fF shunt capacitance at 8~GHz), $Z_{\lambda/4} = 80\ \Omega$, and $Z_{\lambda/2} = 30\ \Omega$. 
Our detailed designing procedure is provided in Appendix~\ref{appendix_network}.
The design principle of any multi-pole PAs can be found in Ref.~\citenum{naaman2022synthesis}.

\section{Kinetic-inductance properties}
\label{sec.NbTiN}

The nonlinearity of kinetic-inductance materials arises from the inertia of Cooper pairs resisting changes in their velocity and becomes more pronounced in thinner films and under current bias.
For films much thinner than their penetration depth, the kinetic inductance is given by
\begin{equation}
    L_\mathrm{k0} = \frac{\hbar}{\pi\Delta_0}\frac{\rho_\mathrm{n}}{d}\frac{l}{w},
\end{equation}
where $\Delta_0$ is the superconducting gap energy, $\rho_\mathrm{n}$ is the normal-state resistivity, $l$, $w$, and $d$ are the KI material's length, width, and thickness, respectively~\cite{zmuidzinas2012superconducting}. 
An injected current enhances $L_\mathrm{k0}$, and we may write an expansion of the current-dependent kinetic inductance~\cite{zmuidzinas2012superconducting},
\begin{equation}
    L_\mathrm{k}(I) = L_\mathrm{k0}\left( 1 + \frac{I^2}{I_\ast^2} + \dotsc \right),
\end{equation}
where $I_\ast$ is the current scale determining the sensitivity~(see Appendix~\ref{appendix_KI} for a detailed discussion).
This is typically approximated as a parabolic function when $I_\ast \gg I$ and written as
\begin{equation}
    L_\mathrm{k}(I) \approx L_\mathrm{k0}\left( 1 + \frac{I^2}{I_\ast^2}\right).\label{eq.Lk_approx}
\end{equation}

Considering a modulated current with a constant offset, $I = I_\mathrm{dc} + I_\mathrm{ac}$, the kinetic inductance becomes 
\begin{equation}
    L_\mathrm{k} \approx L_\mathrm{k0}\left( 1 + \frac{I_\mathrm{dc}^2}{I_\ast^2} + \frac{2I_\mathrm{dc}I_\mathrm{ac}}{I_\ast^2} + \frac{I_\mathrm{ac}^2}{I_\ast^2}\right).
    \label{eq.L_p}
\end{equation}
The fourth term is associated with the Kerr coefficient allowing the four-wave-mixing~(4WM) process, where two pump photons are transformed into signal and idler photons, necessitating $2 \omega_\mathrm{p} = \omega_\mathrm{s} + \omega_\mathrm{i}$, and becomes negligible when $I_\ast \gg |I_\mathrm{ac}|$ as in the case of the NbTiN nanowire we use~(see Appendix~\ref{NbTiN_properties}).

Although some KI-based PAs employ the 4WM process~\cite{bockstiegel2014development,faramarzi20244}, the 3WM process offers advantages in $P_{1\mathrm{dB}}^{\mathrm{in}}$~\cite{frattini2018optimizing,zorin2016josephson} and enables simple pump removal through passive elements~(e.g.\ diplexers) due to a large frequency separation from the signal.
In Fig.~\ref{fig0}(c), a simplified wiring diagram illustrates the paths for the dc current, the signal, and the pump.
With the pump frequency near twice the fundamental frequency of the wavelength $\lambda$, the CPW lines present either half- or full-wavelength to the pump, enabling efficient transmission to the nonlinear inductance with minimal reflections.
A bias tee combines the dc current and the signal frequency before entering the circulator, which redirects the reflected signal into the readout amplification chain.
The diplexer combines the pump and the signal into the three-stage impedance transformer.

In Appendix~\ref{appendix_KI}, we divide $I_\mathrm{ac}$ into a dominant pump current $I_\mathrm{p}$ and a weak signal current represented by a quantum operator, resulting in $\alpha \propto |I_\mathrm{p}|^2$.
Additionally, we approximate that the negative value of the real part of $Y_\mathrm{eff}$ is proportional to $\alpha$ from Eq.~(\ref{eq_eff_Y}), as numerically verified in Appendix~\ref{appendix_simulation}.
Combining the above with the design principle of the three-stage impedance transformer in Appendix~\ref{appendix_network}, we yield the bandwidth relationship,
\begin{equation}
    \Delta\omega \propto \alpha Z_\mathrm{NR} \label{eq_bw_alpha}.
\end{equation}
Any nonlinear resonator, in theory, can utilize a three-stage impedance transformer to widen its bandwidth.
Notably, the bandwidth depends linearly on $\alpha$, implying that the bandwidth constraint is tied to the degree of modulation on the tunable inductors.
For kinetic-inductance materials, the challenge is the finite $\alpha$~[see Eq.~(\ref{eq_xi_limit})] before the onset of superconductivity breakdown, constraining the bandwidth.

\begin{figure}
\centering{}
\includegraphics[width=8.5cm]{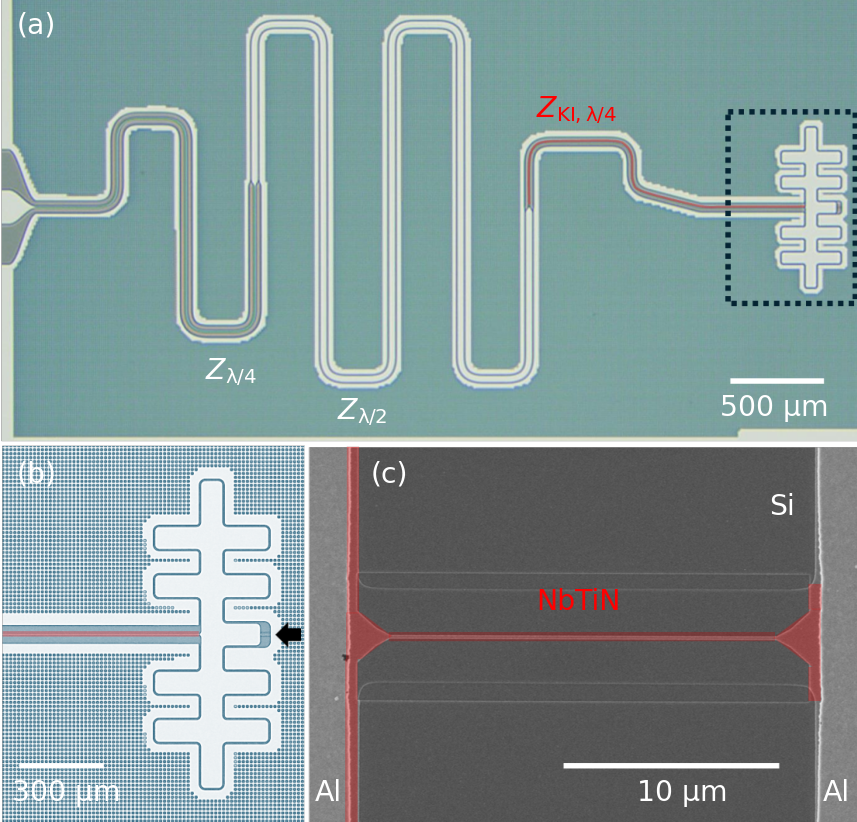}

    \caption{
    (a) Optical images of the device.
    (b)~Magnified image of the dotted rectangular zone in~(a). 
    The high-impedance $\lambda/4$ CPW has a center conductor~(2~$\mu$m in width) made of NbTiN~(red). 
    All other electrodes and ground planes are covered with 120-nm-thick aluminum with low KI.
    There are high-density holes in the ground plane to trap stray vortices.
    The black arrow points at the location of the NbTiN nanowire.
    (c)~Scanning-electron-microscope image showing a 250-nm-wide NbTiN nanowire inductor~(red).}
    \label{fig1}
\end{figure}

\begin{figure*}
\centering{}
\includegraphics[width=14.6cm]{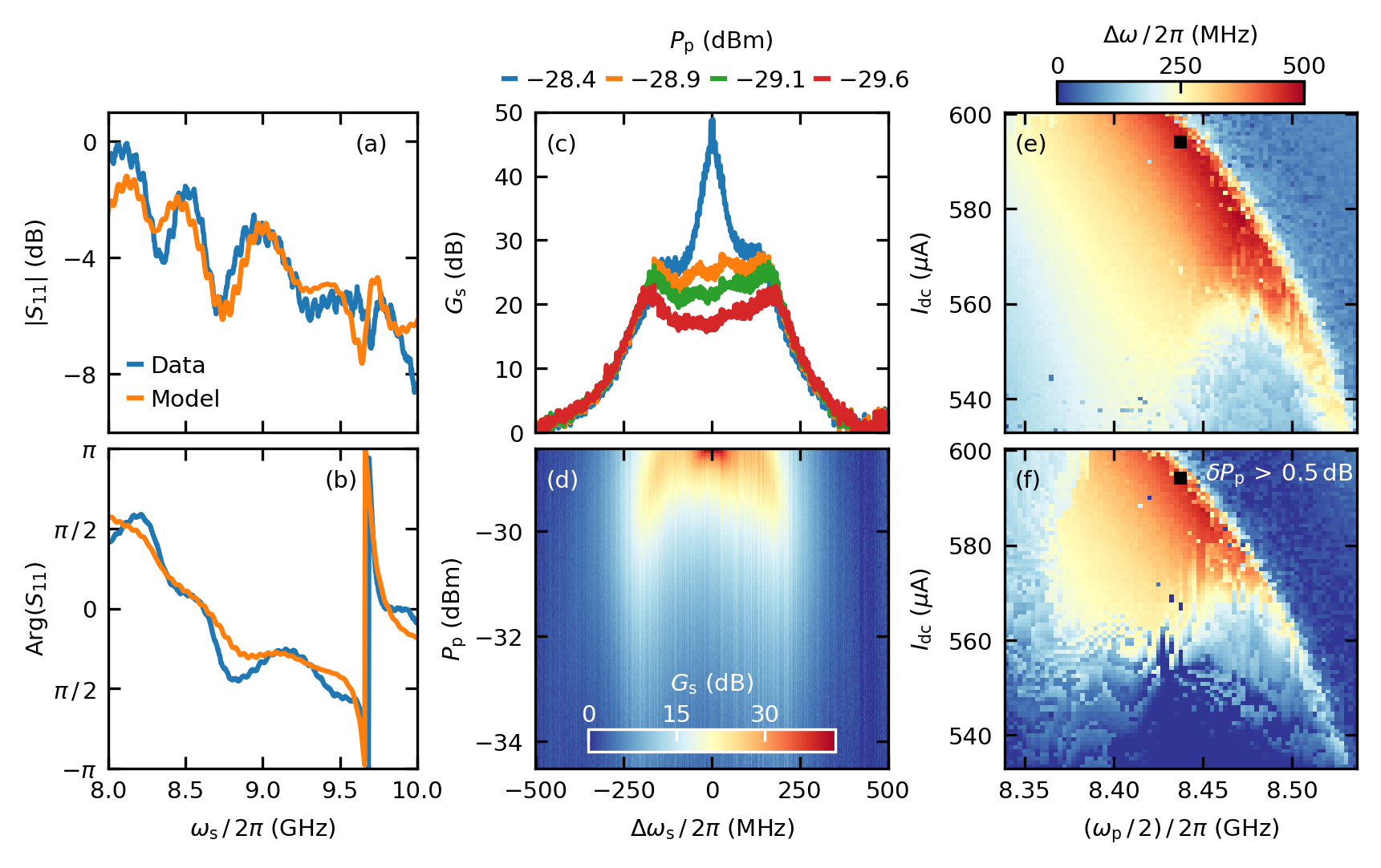}
    \caption{Characterization of the KIMPA.
    (a)~Amplitude and~(b)~phase of the reflection coefficient $S_{11}$ vs signal frequency $\omega_\mathrm{s}$~(blue) and the numerical simulation~(orange).
    (c)~Signal gain $G_\mathrm{s}$ vs signal frequency $\omega_\mathrm{s}$ detuned from half the pump frequency $\omega_\mathrm{p}$ by $\Delta\omega_\mathrm{s} = \omega_\mathrm{s}  -  \omega_\mathrm{p}/2$, at different pump powers $P_\mathrm{p}$.
    (d)~$G_\mathrm{s}$ as a function of $\Delta \omega_\mathrm{s}$ and $P_\mathrm{p}$. 
    (e)~Bandwidth $\Delta \omega$ at 17-dB gain as a function of $\omega_\mathrm{p}$ and $I_\mathrm{dc}$.
    (f)~Refined bandwidth map subject to a stability criterion $\delta P_\mathrm{p} = P_\mathrm{p}^{\mathrm{SC}}/P_\mathrm{p}$ > 0.5~dB to ensure reliable operation. 
    Here, $P_\mathrm{p}^{\mathrm{SC}}$ is the maximum pump power before superconductivity breakdowns occur.
    The breakdown event rate increases further when $P_\mathrm{p}$ approaches $P_\mathrm{p}^{\mathrm{SC}}$.
    The black squares in~(e) and~(f) mark the operating point used in~(c) and~(d), chosen for the optimal balance between bandwidth and gain ripples.}
    \label{fig2}
\end{figure*}

\section{Fabrication Process}
\label{sec_fab}
Our devices incorporate two superconducting materials: high-kinetic-inductance NbTiN and aluminum. 
In Fig.~\ref{fig0}(c), we emphasize two different materials with colors, red for NbTiN and gray for aluminum, in the circuit scheme.
In Figs.~\ref{fig1}(a) and~(b), the NbTiN exposed to air is marked in red, while the remaining electrodes are aluminum-coated on top of NbTiN.
In Appendix~\ref{NbTiN_properties}, we demonstrate that resonators with resonances around 8.5~GHz and nanowire inductors consisting of a thin NbTiN film, constructed through the fabrication methods detailed below, exhibit internal quality factors $Q_\mathrm{i}\approx10^5$.

Fabrication begins with cleaning a 3-inch 300-$\mu$m-thick high-resistivity silicon wafer using hydrofluoric acid~(HF) to remove surface oxide, followed by sputtering of NbTiN at room temperature. 
We then pattern nanowire inductors using electron-beam lithography with ZEP520A resist and etch the exposed and developed regions using CF$_4$ dry etching. 
Subsequently, we define the CPWs, PAs, and vortex traps through photolithography using AZ1500-4.4cp resist and CF$_4$ dry etching.
After another HF cleaning step, we perform ion-milling surface cleaning under a vacuum before depositing aluminum of 150~nm.
The aluminum layer is then patterned with photolithography and wet etched to cover all electrodes except high-impedance $\lambda/4$ CPWs and nanowire inductors.
This Al-on-NbTiN bilayer serves a few purposes:~(i)~ensuring low kinetic inductance to match our COMSOL simulations, (ii)~reducing the radiation loss, and (iii)~reducing stray inductance.
Without aluminum, two CPW sections would require CPW center widths $w$ several times larger than our current design, leading to extra radiation loss $1/Q_{\mathrm{rad}} \propto w^2$~\cite{mazin2005microwave,zmuidzinas2012superconducting}.
Furthermore, increased stray inductance would dilute the nonlinearity and effectively enhance $I_\ast$, requiring more pump power.
Finally, the wafer is coated with a protective photoresist layer, diced into 2.5$\times$5~mm$^2$ chips, and cleaned with N-Methylpyrrolidone remover.

\section{Device performance}
\label{sec_measurement}
\subsection{Gain measurement}
The devices are measured in a dilution refrigerator at a base temperature of 10~mK.
The amplitude and phase of the reflection coefficient $S_{11}$ are shown in Figs.~\ref{fig2}(a) and~(b).
The NR resonance frequency in the absence of $I_\mathrm{dc}$ is approximately 9.6~GHz.
The measurement curve in blue exhibits noticeable ripples arising from impedance mismatches at the wire bonding, circulators, diplexers, etc.
The distance between impedance mismatches creates standing waves that affect the gain profile through constructive or destructive interference depending on the frequency.
The experimental setup significantly influences the gain profile even for identical devices, which is a phenomenon noted in previous studies~\cite{mutus2014strong,roy2015broadband,ranzani2022wideband}.
To minimize interference effects, we directly connect the diplexer and circulator, eliminating interconnecting cables.

The non-ideal environment is modeled as two superimposed sinusoidal waves on top of the 50-$\Omega$ baseline: 
\begin{equation}
    Z_{\mathrm{env}}(\omega)=Z_0 + Z_1e^{i(\omega \tau_1 + \phi_1)} + Z_2e^{i(\omega\tau_2 + \phi_2)},\label{eq.z_env}
\end{equation}
where $Z_0$ = 50~$\Omega$, $Z_n$~($n$=1,2) is the amplitude of periodic modulations, $\tau_n$ is the period, and $\phi_n$ is the phase offset.
The orange curve shows our numerical calculation with parameters $Z_1$ = 14.2~$\Omega$, $Z_2$ = 1.9~$\Omega$, $\tau_1/2\pi$ = $10.5$~ns, $\phi_1$ = $-0.7\pi$, and $\tau_2/2\pi$ = $121$~ns, and $\phi_2$ = 0 to match the ripple of a period of $\sim$450~MHz and a 4-dB amplitude.
Interestingly, $\tau_1$ corresponds to approximately a 17-cm coaxial cable with a phase velocity of $2\times10^8$~m/s, closely matching the device-to-circulator distance.

To characterize the amplifier's performance, we apply $I_\mathrm{dc}$ ranging from 530 to 600~$\mu$A to tune the NR close to 8.47~GHz. 
The pump frequency $\omega_\mathrm{p}$ is set to approximately twice the resonance frequency, with the pump power $P_\mathrm{p}$ gradually increasing until reaching the superconductivity-breaking point $P_\mathrm{p}^{\mathrm{SC}}$.
An example of the measured signal gain $G_\mathrm{s}$ vs $\Delta\omega_\mathrm{s} = \omega_\mathrm{s}  - \omega_\mathrm{p}/2$ at different $P_\mathrm{p}$ is shown in Figs.~\ref{fig2}(c) and~(d).
The amplifier exhibits a 450-MHz bandwidth with a signal gain exceeding 17~dB at $P_\mathrm{p} = -29.6$~dBm.

We observe that the curves have ripples in the valley between two peaks due to the impedance mismatch, which become more pronounced at higher pump powers and higher gain. 
In an ideal 50-$\Omega$ matched environment, we expect two local maxima, but four peaks are ultimately observed.  
The four peaks can be quantitatively reproduced using  $Z_\mathrm{env}$ in Eq.~(\ref{eq.z_env}) as depicted in Fig.~\ref{sfig_simulation} in Appendix~\ref{appendix_simulation}.  
A center peak grows with $P_\mathrm{p}$, and eventually $G_\mathrm{s}$ exceeds 40~dB, but this comes at the cost of the reduced gain--bandwidth product, whose behavior is captured also in the simulation in Appendix~\ref{appendix_network}.
When $P_\mathrm{p}\ >\ P_\mathrm{p}^{\mathrm{SC}}$ = $-$28.4~dBm, the superconductivity breaks down, resulting in complete loss of amplification.
Recovering the superconducting state requires turning off both the pump and dc current.

Fig.~\ref{fig2}(e) reveals a maximum bandwidth of 470~MHz.
However, as $P_\mathrm{p}$ approaches $P_\mathrm{p}^{\mathrm{SC}}$, there is an increase in the stochastic superconductivity breaking events, likely because $I_\mathrm{dc} + I_\mathrm{ac}\rightarrow I_\mathrm{c}$ and a higher probability of pair-breaking photon absorption~\cite{mazin2009microwave}.
Since there is no obvious difference in the event rate of superconductivity breakdown~(occurring every 2--6 hours) when $\delta P_\mathrm{p} = P_\mathrm{p}^{\mathrm{SC}}  / P_\mathrm{p}>$ 0.5~dB, we set the threshold to ensure reliable operation.
The revised plot is in Fig.~\ref{fig2}(f).
The black square indicates the operating parameters in Figs.~\ref{fig2}(c) and~(d), which lead to less ripple amplitude even though the bandwidth is smaller than the maximum achievable bandwidth.
In the simulation, we demonstrate that the maximal available bandwidth under $Z_
\mathrm{env}$ is down to about 500-MHz from the designed 600-MHz, even though the impedance mismatch can occasionally help in smoothing gain ripples at certain $\omega_\mathrm{p}$ and $I_\mathrm{dc}$.
Enhancing impedance matching is crucial to unlock the potential of the three-stage impedance transformer.

\subsection{Saturation-power characteristics}
\begin{figure}
\centering{}
\includegraphics[width=8.5cm]{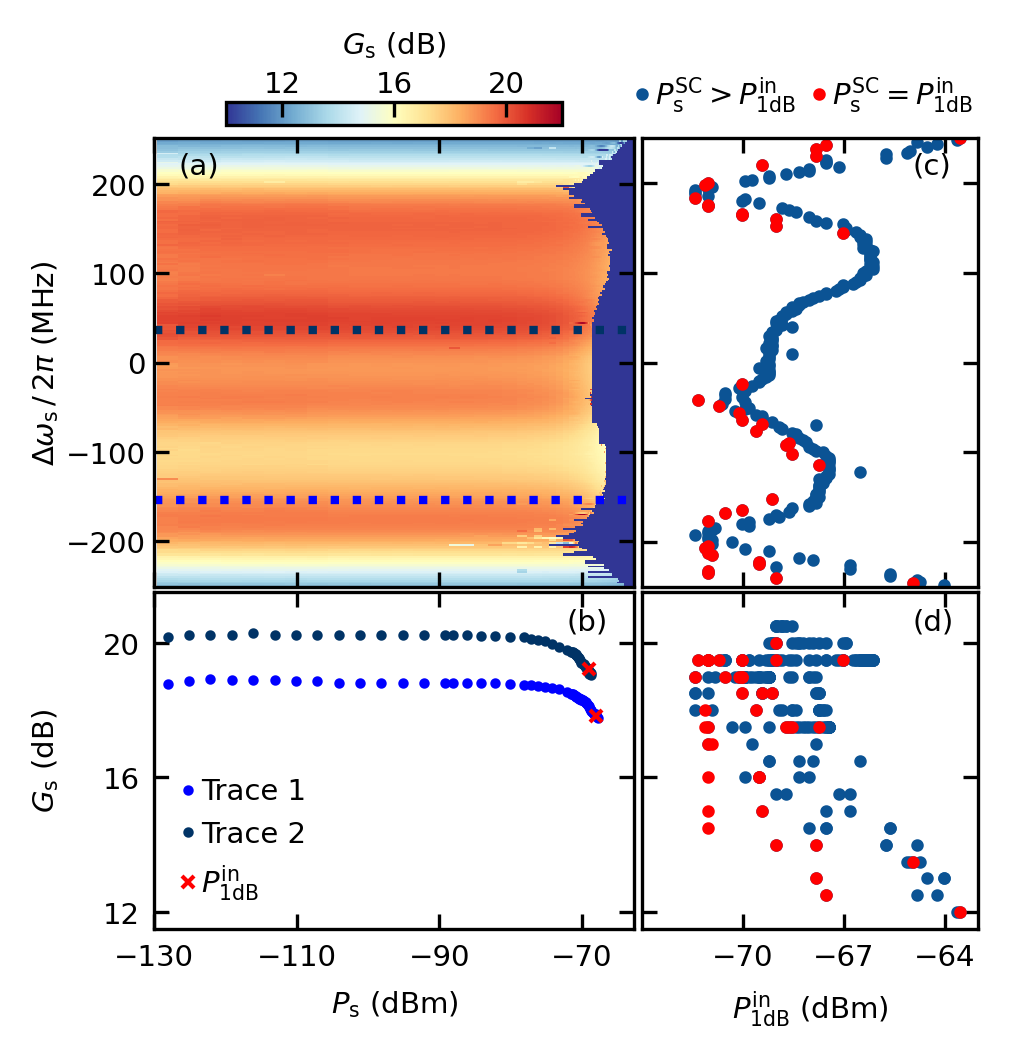}
    \caption{
    Saturation-power characterization of the KIMPA.
    (a)~Color map showing the signal gain $G_\mathrm{s}$ as a function of $\Delta\omega_\mathrm{s}$ and the signal power $P_\mathrm{s}$.
    $G_\mathrm{s}$ shows an abrupt drop-off at high $P_\mathrm{s}$ rather than gradual compression on the right side, indicating the breakdown of superconductivity, where this power level is denoted as $P_\mathrm{s}^\mathrm{SC}$.
    $P_\mathrm{s}^\mathrm{SC}$ is sometimes smaller than the input 1-dB compression power, $P_{1\mathrm{dB}}^{\mathrm{in}}$, and thus we define $P_{1\mathrm{dB}}^{\mathrm{in}}$ = $P_\mathrm{s}^\mathrm{SC}$ in such cases.
    (b)~Examples of gain compression curves for two different operating points, with red crosses indicating $P_{1\mathrm{dB}}^{\mathrm{in}}$. The data correspond to those on the dashed lines in~(a). 
    (c)~$P_{1\mathrm{dB}}^{\mathrm{in}}$ as a function of $\Delta\omega_\mathrm{s}$. We mark red when $P_{1\mathrm{dB}}^{\mathrm{in}}$ = $P_\mathrm{s}^\mathrm{SC}$.
    (d)~Scatter plot of $P_{1\mathrm{dB}}^{\mathrm{in}}$ versus the corresponding $G_\mathrm{s}$ in 0.5-dB increments, indicating variations in $P_{1\mathrm{dB}}^{\mathrm{in}}$ even at constant gain levels.}
    \label{fig3}
\end{figure}

In Fig.~\ref{fig3}(a), we characterize $P_{1\mathrm{dB}}^{\mathrm{in}}$ by incrementing $P_\mathrm{s}$ using the same parameters as used at the black square in Fig.~\ref{fig2}(e)  with $P_\mathrm{p}$ = $-$29.7~dBm.
The color map in Fig.~\ref{fig3}(a) illustrates $G_\mathrm{s}$ as a function of $\Delta\omega_\mathrm{s}$ and $P_\mathrm{s}$.
Rather than exhibiting gradual compression, the gain drops off abruptly, indicating superconductivity breakdown. 
We denote this critical power level as $P_\mathrm{s}^\mathrm{SC}$.

Two examples of the cross-section, $P_\mathrm{s}$ vs $G_\mathrm{s}$, are shown in Fig.~\ref{fig3}(b).
Red crosses indicate $P_{1\mathrm{dB}}^{\mathrm{in}}$.
At some frequencies $P_\mathrm{s}^\mathrm{SC}$ occurs before reaching the conventional 1\nobreakdash-dB compression point, effectively making $P_{1\mathrm{dB}}^{\mathrm{in}} = P_\mathrm{s}^\mathrm{SC}$ in those cases.
We plot $P_{1\mathrm{dB}}^{\mathrm{in}}$ vs $\Delta\omega_\mathrm{s}$ in Fig.~\ref{fig3}(c) and highlight data points in red when $P_{1\mathrm{dB}}^{\mathrm{in}}$ = $P_\mathrm{s}^\mathrm{SC}$.
In Fig.~\ref{fig3}(d), we have both $G_\mathrm{s}$ and $P_{1\mathrm{dB}}^{\mathrm{in}}$ in 0.5\nobreakdash-dB increments and observe variations in $P_{1\mathrm{dB}}^{\mathrm{in}}$ even at constant gain levels, which we attribute to stochastic superconductivity-breaking events and the interference with standing-wave modes in the environment.

The output saturation power $P_{1\mathrm{dB}}^{\mathrm{out}}$ is defined as the output amplified signal power $P_\mathrm{s}$ at which $G_\mathrm{s}$ decreases by 1~dB compared to its weak-power limit.
Compared to other single-pole KI PAs in Ref.~\citenum{parker2022degenerate}, our $P_{1\mathrm{dB}}^{\mathrm{out}}$ ranges between $-$55 and $-$49~dBm at 17-dB~gain similar to their $-$50~dBm.
However, we have approximately tenfold bandwidth $\Delta\omega$ expecting a 20-dB increase in $P_{1\mathrm{dB}}^{\mathrm{out}}\propto\Delta\omega^2$~\cite{zorin2016josephson,frattini2018optimizing}.
This reduction might be attributed to an enhanced Kerr coefficient, which leads to an ac Stark shift, caused by the small cross-section nanowire and higher-order effects induced by a much higher ratio of $|I_\mathrm{p}|$ over the critical current $I_\mathrm{c}$ than others~\cite{clem2012kinetic}.

\subsection{Noise characteristics}

\begin{figure}
\centering{}
\includegraphics[width=7.15cm]{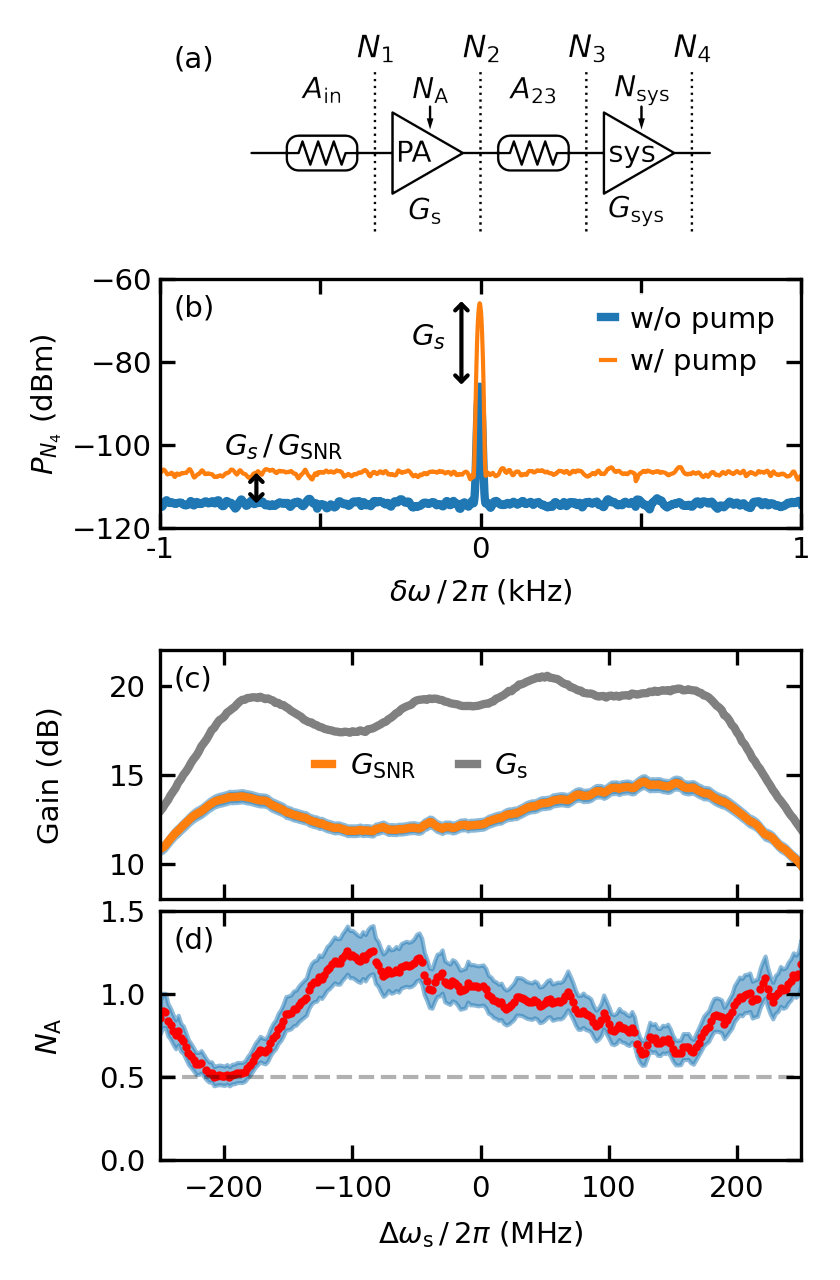}
    \caption{
    (a)~Model for calibrating the added noise $N_\mathrm{A}$. We consider loss components,  $A_\mathrm{in}$ and $A_{23}$, the KIMPA gain~$G_\mathrm{s}$ and the subsequent gain~$G_{\mathrm{sys}}$ in the measurement chain.
    (b)~Noise spectrum obtained with a spectrum analyzer.
    The peak difference with and without the pump represents the KIMPA gain.
    Similarly, the signal-to-noise-ratio~(SNR) gain $G_{\mathrm{SNR}}$ is indicated by the noise-floor difference as displayed in the panel.
    (c)~$G_{\mathrm{SNR}}$~(orange) and $G_\mathrm{s}$~(gray) as a function of the signal-frequency detuning~$\Delta\omega_\mathrm{s}$.
    (d)~Added noise $N_\mathrm{A}$~(red) as a function of $\Delta\omega_\mathrm{s}$. $N_\mathrm{A}$ is extracted using Eq.~(\ref{eq.added_noise}). 
    Error bars~(blue) are determined by the fitting inaccuracy and the uncertainty in the microwave component difference between the qubit and KIMPA measurement chains~(see Appendix~\ref{appen_sys_noise}).
    The gray dashed line represents the quantum-limited noise of 0.5 quanta.
    }
    \label{fig4}
\end{figure}

The added noise $N_\mathrm{A}$ of PAs determines the noise performance and is lower-bounded by the uncertainty principle of quantum mechanics to 0.5~quanta in the case of non-degenerate amplifications~\cite{PhysRevD.26.1817}.
The circuit diagram, as depicted in Fig.~\ref{fig4}(a), is used to model the noise added by different components.
The cascaded noise equations during pumping are written as follows:
\begin{align}
    N_2 &= G_\mathrm{s}(N_1 + N_\mathrm{A}), \\
    N_3 &= A_{23}N_2 + (1-A_{23}) N_{T_{23}}, \\
    N_4 &= G_{\mathrm{sys}}(N_3 + N_{\mathrm{sys}}).
\end{align}
Here, $N_i$~($i = 1,2,3,4$) is the noise at each stage, $A_\mathrm{in}$ is the loss between the input source and the PA, $A_{23}$ is the loss between the PA and the subsequent amplifiers~(HEMT plus the room temperature amplifier), $N_{T_{23}}$ is the extra noise from the environment of the temperature $T_{23}$ , $G_{\mathrm{sys}}$ is the gain of the subsequent amplifier, and $N_{\mathrm{sys}}$ is the effective added noise.
With sufficient attenuation and infrared filtering on the input line~($A_\mathrm{in}<-80$~dB, in this instance), we anticipate the actual temperature of electrons of our device is reduced down to below 100~mK, which yields $N_1 = \frac{1}{2}\coth{(\hbar\omega/k_\mathrm{B}T)}\approx0.5$~\cite{callen1951irreversibility,kerr1997receiver}.

In the case without pumping, we append ``$\mathrm{off}$'' to $N_i$ and assign $N_2^{\mathrm{off}} \approx0.5$ because the PA reflects the signal without adding noise, while the second and third equations remain unchanged.
The final-stage noise $N_4 = P_{N_4} / (\hbar\omega B_\mathrm{m})$ is quantified by a spectrum analyzer, where $P_{N_4}$ is the measured background noise floor and $B_\mathrm{m}=10$~Hz denotes the IF bandwidth used for the measurement.
Finally, by subtracting $N_4^{\mathrm{off}}$ from $N_4$, we obtain 
\begin{equation}
    N_\mathrm{A} = \frac{N_4  - N_4^{\mathrm{off}}}{G_\mathrm{s}A_{23}G_{\mathrm{sys}}} + \frac{N_1}{G_\mathrm{s}}  - N_1. \label{eq.added_noise}
\end{equation}

In Fig.~\ref{fig4}(b), we show spectra obtained from the spectrum-analyzer measurements with the pump on~(orange) and off~(blue), respectively.
The gain in the signal-to-noise ratio $G_{\mathrm{SNR}}$ can be obtained from the shift of the noise floor, where $P_{N_4} / P_{N_4}^{\mathrm{off}} = G_\mathrm{s} / G_{\mathrm{SNR}}$.
$G_{\mathrm{SNR}}$ vs $\Delta\omega_\mathrm{s}$ is plotted in Fig.~\ref{fig4}(c) compared to $G_\mathrm{s}$ in gray.
We observe an approximate 5-dB difference between $G_{\mathrm{SNR}}$ and $G_\mathrm{s}$. 

All variables in Eq.~(\ref{eq.added_noise}) can be extracted from spectrum-analyzer measurements, except for the effective system gain $G_\mathrm{sys}^\mathrm{eff} = A_{23}G_{\mathrm{sys}}$ calibrated using a qubit~\cite{mirhosseini2019cavity,qiu2023broadband}.
We use a microwave switch to transmit the signal through a second input line, where a waveguide is coupled with a tunable transmon qubit, and measure the saturation of the transmission coefficient $S_{21}$ of the waveguide while increasing the signal power.
In Appendix~\ref{appen_sys_noise}, we describe the method to obtain $A_\mathrm{in}$ and $G_\mathrm{sys}^\mathrm{eff}$. 

In Fig.~\ref{fig4}(d), the PA shows a near quantum-limited added noise with $0.5 < N_\mathrm{A} < 1.3$ in the bandwidth.
The extra added noise $N_\mathrm{ex}$ besides the quantum-limited noise~(0.5~quanta) is written as~\cite{parker2022degenerate}
\begin{align}
    &N_\mathrm{A}  = 0.5 + N_\mathrm{ex}, \\
    &N_\mathrm{ex} = \frac{Q_\mathrm{e}}{2Q_\mathrm{i}}\frac{(\sqrt{G_\mathrm{s}} + 1)^2 }{G_\mathrm{s}-1}\left( 2 N_\mathrm{th} + 1\right) + N_\mathrm{th}, \\
    &N_\mathrm{th} = \frac{1}{e^{\frac{\hbar\omega}{k_\mathrm{B}T}}-1},
\end{align}
where $N_\mathrm{th}$ is the thermal occupation photon number and $Q_\mathrm{e}$ is the external quality factor of the PA.
A coaxial cable with an attenuation constant $\alpha$ = 0.5~dB/m has an internal quality factor $Q_\mathrm{i} = \pi/{(\lambda\alpha)}= 3410$~\cite{pozar2021microwave,kurpiers2017characterizing} and the wavelength $\lambda$ of 34~cm, while lossy elements~(e.g. diplexer and microwave connectors) can further reduce $Q_\mathrm{i}$.
In Fig.~\ref{fig4}(c), we expect two local maxima in our two-pole system, but four are observed in $G_{\mathrm{s}}$. In contrast, only two local maxima appear in $G_{\mathrm{SNR}}$ suggesting that the coupled cable modes are likely the dominant noise source.

Furthermore, the strong pump raised the base temperature from 10~mK to 25~mK primarily due to dissipation in the 1-dB attenuator and the Eccosorb filter before the diplexer~(see our setup in Fig.~\ref{sfig_setup}). 
This heating results in additional thermal noise that is transmitted through the diplexer into the device in a frequency-dependent manner due to impedance mismatch, increasing the added noise.
Nevertheless, we suspect that the increased dielectric loss, due to fields extending into lossy dielectrics in the coaxial cables, plays a more crucial role than thermal injection, since thermal noise at the device frequency should be significantly diminished by the diplexer’s high-pass filtering characteristics.
More efforts are needed to minimize the standing wave effect and to verify the origins of $N_\mathrm{ex}$.

\section{Conclusion and PROSPECTS}
In this work, we have demonstrated the kinetic-inductance impedance-engineered parametric amplifier~(KIMPA).
Our three-stage impedance-transformer scheme successfully extended the range of impedance values for the nonlinear resonator up to several tens of ohms while reducing shunt-capacitance requirements from several picofarads to 330~fF.
Using 20\nobreakdash-nm-thick, 250\nobreakdash-nm-wide NbTiN nanowires as the nonlinear element, the device achieved a 450-MHz bandwidth at 17-dB gain and provides approximately $-$51-dBm output saturation power and near quantum-limited noise performance of 0.5--1.3~quanta of added noise.
Our circuit shines a light on other three-wave-mixing building blocks to implement broad-bandwidth parametric amplifiers allowing high dynamic range or operation in strong magnetic fields.

The aluminum coating on NbTiN as described in Sec.~\ref{sec_fab}, while beneficial for millikelvin operation, prevents high-temperature or high-magnetic-field operation.
Replacing aluminum with a thick niobium film may enable KIMPAs to function in environments beyond millikelvin temperatures without significantly altering the current design. 
Alternatively, a fully kinetic-inductance implementation of the three-stage transformer, feasible through precise length adjustments of the CPWs to maintain quarter- and half-wavelength matching, may also enable KIMPA functionality in varied conditions.  

Future developments could address the two main issues of impedance mismatches and insufficient quasiparticle protection.
The impedance mismatch brings uncertainty in the bandwidth relative to the simulation, potentially decreasing the bandwidth depending on the phase offset of sinusoidal waves in $Z_\mathrm{env}$ according to the simulation performed in Appendix~\ref{appendix_simulation}.
The existing diplexer, which undergoes impedance changes from room temperature to millikelvin temperature, occupies a substantial 10-cm footprint, hindering our ability to reduce the distance to the circulator.
Quasiparticles generated by the absorption of cosmic rays could break down the superconductivity especially when the signal power is near $P_\mathrm{s}^\mathrm{SC}$. 
Moreover, quasiparticles from the pump line persist because of the insufficient pump power from the room-temperature microwave source leading to insufficient attenuation. 

To tackle these two challenges, we propose integrating KIMPA with on-chip components, for example, on-chip diplexers and on-chip filters. 
On-chip elements enable better impedance matching, decrease the coupling to standing wave modes and, in optimal conditions, may slightly improve bandwidth.
Moreover, a smaller physical footprint permits complete shielding of the KIMPA under a $\mu$-metal cover, shielding it from magnetic noise. 
Additionally, on-chip components enhance pump-power efficiency and noise isolation from the pump port \cite{dai2024optimizing}, potentially minimizing attenuations at the millikelvin stage to reduce heat generation.

The inherent properties of NbTiN impose a maximal modulation strength $\alpha$, thereby limiting the attainable bandwidth as outlined in Eq.~(\ref{eq_bw_alpha}). 
According to Eq.~(\ref{eq_xi_limit}), our KIMPA may approach this $\alpha$-limit when $I_\mathrm{dc}$ is roughly half of $I_\ast$. 
To further enhance the bandwidth, aside from reducing impedance mismatch, initiating with a smaller nonlinear resonator $Z_\mathrm{NR}$ that exhibits better pump efficiency, as detailed in Appendix~\ref{Appedix_roy}, and subsequently refining the three-stage impedance transformer could be beneficial. 
Alternatively, reducing $I_\mathrm{dc}$, potentially to one third of $I_\ast$, while increasing $I_\mathrm{p}$, can slightly raise $\alpha$ due to the higher-order nonlinear effect from $I_\mathrm{p}$, which is not captured in Eq.~(\ref{eq.xi}), but may also diminish $P_{1\mathrm{dB}}^{\mathrm{in}}$ from ac Stark shifting the resonance of the nonlinear resonator~\cite{frattini2018optimizing}.

Besides using kinetic-inductance materials, interest is increasing in using alternative 3WM components for PAs, such as graphene JJs~\cite{butseraen2022gate,sarkar2022quantum} and InAs nanowires~\cite{splitthoff2024gate}, operating under various conditions that conventional JJs cannot.
These nonlinear inductors exhibit approximately 500 pH, equivalent to $Z_{\mathrm{NR}}$ = 25~$\Omega$ in an LC resonator at 8~GHz, hindering implementations in the conventional circuit.
Moreover, our circuit can advantageously be applied to JJ-based PAs. 
Increasing the number of serial JJ units, $M$, could improve $P_{1\mathrm{dB}}^{\mathrm{in}}$ (which scales with $M$)~\cite{frattini2018optimizing} but also increases $Z_{\mathrm{NR}}$~(which scales with $\sqrt{M}$). While JJ-based PAs with $M\ =\ 67$ exist~\cite{ezenkova2022broadband} using a conventional circuit, our circuit allows increasing $M$ to 1000, which has shown output saturation power of $-$88~dBm in a single-pole PA~\cite{sivak2020josephson} and is expected to improve $P_{1\mathrm{dB}}^{\mathrm{in}}$ by more than 10~dB using our circuit due to the external-coupling-rate improvement~\cite{frattini2018optimizing}.

\section{Acknowledgment}
We acknowledge Y. Hishida and H. Terai at the National Institute of Information and Communications Technology for the NbTiN-film deposition.
The work was supported in part by the Ministry of Education, Culture, Sports, Science and Technology~(MEXT) Quantum Leap Flagship Program~(QLEAP)
(Grant No.~JPMXS0118068682), the JSPS Grant-in-Aid for Scientific Research~(KAKENHI)~(Grant No.~JP22H04937), and the JSPS JRP-LEAD with UKRI~(Grant No.~JSPSJRP20241713).

\setlength{\tabcolsep}{7pt}
\begin{table*}[t]  
\caption{Overview of kinetic-inductance parametric amplifiers. The table compares kinetic-inductance material, operating conditions, and performance metrics.
Magnetic-field orientations are indicated as perpendicular~($\perp$) or parallel~( $\mathrel{\!/\!\!/\!}$ ) to the device plane, with ``--'' indicating no applied field.
$I_\ast$ represents the current sensitivity, WM denotes wave-mixing order~(3 for three-wave mixing, 4 for four-wave mixing), GBWP is the gain--bandwidth product, and $P_\mathrm{1dB}^\mathrm{out}$ is the output saturation power at 1-dB compression.
$N_\mathrm{A}$ shows the added noise in photon number units, with ``n/a'' indicating unreported values.}
\begin{tabular}{lcccccccc}
\toprule
Author & Material & 
\parbox[c]{1.6cm}{\centering Temperature\\(mK)} & 
\parbox[c]{1.2cm}{\centering Field\\(Tesla)} &
\parbox[c]{1.0cm}{\centering $I_\ast$\\(mA)} &
\parbox[c]{1.2cm}{\centering WM} &
\parbox[c]{1.4cm}{\centering GBWP\\(MHz)} &
\parbox[c]{1.2cm}{\centering $P_\mathrm{1dB}^\mathrm{out}$\\(dBm)} &
\parbox[c]{1.2cm}{\centering $N_\mathrm{A}$\\(quanta)}\\
\midrule
Parker \etal~\cite{parker2022degenerate} & NbTiN & 20 &-- & 5.1 & 3 & 54 & $-$49 & 0.7\footnote{Estimation from the phase-sensitive added noise.}\\
Xu \etal~\cite{xu2023magnetic} & NbN & 8 & 0.5 ($\perp$) & -- & 4 & 74 & $-$109 & $< 0.6$ \\
Khalifa \etal~\cite{khalifa2023nonlinearity} & NbTiN & 8 & 2 ($\perp$) & -- & 4 & 5 & $-$104 & n/a \\
Vaartjes \etal~\cite{vaartjes2024strong} & NbTiN & 1800 & 2 ($\perp$) & 5.1 & 3 & 17 & $-$50 & $< 0.6$ \\

Khalifa \etal~\cite{khalifa2024kinetic} & NbTiN & 10 & -- & $\sim$1 & 3 & 45 & $-$73 & 2.6 \\
Splitthoff \etal~\cite{splitthoff2024gate} & NbTiN & 30 & 0.5 ( $\mathrel{\!/\!\!/\!}$ ) & -- & 4 & 30 & $-$100 & n/a\footnote{Near quantum-limited noise.}\\
Zapata \etal~\cite{zapata2024granular} & GrAl & 30 & 1 ($\perp$) & -- & 4 & 28 & $-$90 & $0.5\text{--}1.2$ \\
Frasca \etal~\cite{frasca2024three} & NbN & 850 & 6 ($\perp$) & 0.35 & 3 & 20 & $-$65 & $\sim$0.5 \\
Mohamed \etal~\cite{mohamed2024selective} & NbTiN & 4500 & -- & 5.86 & 3 & 30 & $-$53 & 1.3 \\
This work & NbTiN & 10 & -- & 1.65 & 3 & 3200 & $-$51 & 0.5--1.3\\
\bottomrule
\end{tabular}\label{table.KPA}
\end{table*}
\setlength{\tabcolsep}{3.6pt}

\begin{table*}[t]  
\caption{Overview of broadband parametric amplifiers. The table compares different nonlinear element (NLE) implementations and their performance characteristics.
$f_\mathrm{r}$ is the center frequency, BW is the bandwidth, and Norm.~BW is the normalized bandwidth $\mathrm{BW}$/$f_\mathrm{r}$. $C$ represents the capacitance, and $Z_\mathrm{NR}$ is the impedance of the nonlinear resonator.
Method indicates the broadening approach: extra modes from TL~(transmission line) and LE~(lumped-element) resonators, or exhanced coupling from KT~(Klopfenstein taper) and Ruthroff transformer.}
\setcounter{footnote}{2}
\begin{tabular}{lcccccccccc}
\toprule
Author & NLE & 
\parbox[c]{1.0cm}{\centering Gain\\(dB)} & 
\parbox[c]{0.8cm}{\centering $f_\mathrm{r}$\\(GHz)} &
\parbox[c]{1.0cm}{\centering BW\\(MHz)} &
\parbox[c]{1.0cm}{\centering Norm. BW} &
\parbox[c]{0.7cm}{\centering $C$ \\(pF)} &
\parbox[c]{0.7cm}{\centering $Z_\mathrm{NR}$\\($\Omega$)} &
\parbox[c]{0.8cm}{\centering Method} &
\parbox[c]{1.2cm}{\centering $P_\mathrm{1dB}^\mathrm{out}$\\(dBm)} &
\parbox[c]{1.2cm}{\centering $N_\mathrm{A}$\\(quanta)}\\
\midrule
Roy \etal~\cite{roy2015broadband} & SQUID & 20 & 5.9 & 640 & 11\% & 3.4 & 7.9 & 2-pole (TL) & $-$90 & $\sim$0.5\\
Grebel \etal~\cite{grebel2021flux} & SQUID & 20--25 & 5.3 & 300 & 5.7\% & 2 & 6 & 2-pole (TL) & $-$96 & $\sim$0.5\\
Ezenkova \etal~\cite{ezenkova2022broadband}\footnote{Series LC circuit. Not accommodate dc-current injection.} & SNAIL arr. & 17 & 6.4 & 300 & 6.4\% & 0.03 & 110 & 2-pole (TL) & $-$80 & $\sim$0.5\\
Kaufman \etal~\cite{kaufman2023josephson} & rf-SQUID arr. & 20 & 4.9 & 500 & 10.2\% & 6.6 & 4.9 & 3-pole (TL + LE) & $-$73 & n/a\footnote{Near quantum-limited noise.} \\
This work & NbTiN & 17 & 8.4 & 450 & 5.2\% & 0.3 & 56 & 2-pole (TL) & $-$51 & 0.5--1.3\\

\midrule
Mutus \etal~\cite{mutus2014strong} & SQUID & 15--23 & 6.7 & 700 & 10.5\% & 4.8 & 5 & KT & $-$95 & $\sim$0.5\\
Lu \etal~\cite{lu2022broadband} & SQUID & 20 & 6.6 & 250 & 3.8\% & 4.9 & 5 & KT & $-$95 & $\sim$0.5\\
White \etal~\cite{white2023readout} & rf-SQUID arr. & $>$20 & 4.7 & 300 & 6.4\% & 6.5 & 5.2 & KT & $-$75 & $\sim$0.8\\
Qing \etal~\cite{qing2024broadband} & SQUID & 20 & 6.6 & 200 & 3\% & 4 & 6 & KT & $-$90 & 0.5--0.9\\
\midrule
Ranzani \etal~\cite{ranzani2022wideband} & SQUID & 17 & 6.3 & 450 & 7.1\% & -- & -- & Ruthroff & $-$105 & 0.5--1.1\\
\bottomrule
\end{tabular}\label{table_BPA}
\end{table*}

\setcounter{section}{0}
\appendix

\section{Comparison to other implementations}
\label{appen_table}

To contextualize this work within the broader landscape of parametric amplifiers, we present two tables that compile performance metrics from single-pole kinetic-inductance and broadband Josephson-junction parametric amplifiers across numerous recent publications. 

Table~\ref{table.KPA} examines kinetic-inductance parametric amplifiers utilizing various superconducting materials including NbTiN, NbN, and granular aluminum~(GrAl). Operating conditions span a wide range, from millikelvin to several-kelvin temperatures, with magnetic field configurations varying from in-plane to perpendicular orientations or field-free operation. The wave-mixing processes predominantly employ three-wave mixing~(3WM) or four-wave mixing~(4WM), with gain--bandwidth products typically ranging from 5 to 74 MHz in previous works.

These implementations demonstrate only in nonlinear elements related to Josephson junctions, ranging from single superconducting quantum-interference-devices~(SQUIDs) to more complex architectures including SQUID arrays, rf-SQUID arrays, and superconducting nonlinear asymmetric inductive element~(SNAIL) arrays.
The strategies reflect the ongoing evolution of two impedance-matching techniques,~(i)~including adding extra modes through transmission line or lumped-element resonators,~(ii)~utilizing Klopfenstein tapers or Ruthroff transformers for smooth impedance transitions. 
Note that nonlinear resonators in all implementations employ a parallel LC circuit except Ref.~\citenum{ezenkova2022broadband}, which has a series LC resonator and does not accommodate dc-current injection for our application.

\section{Analysis of a synthesized impedance-matched network}
\subsection{Parametric amplification with a modulated inductor}
\label{appendix_modulate_inductor}
\begin{figure}
    \centering
    \includegraphics[width = 5.5cm]{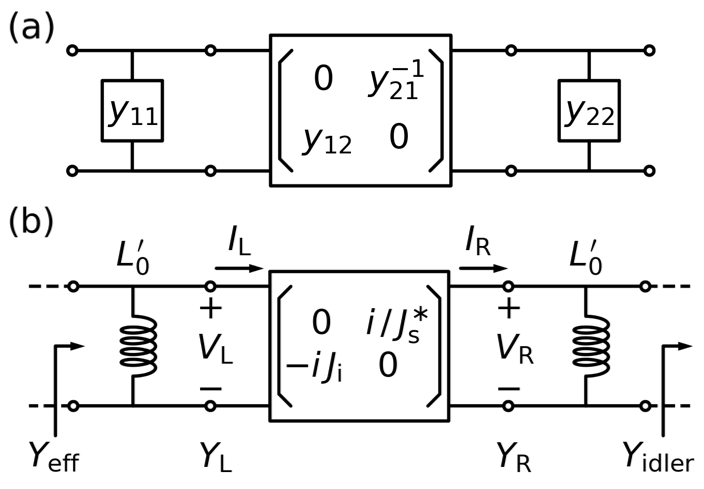}
    \caption{Circuit diagram of a parametrically modulated inductor~[Eq.~(\ref{eq_abcd})].
    (a)~Illustration using admittance matrix parameters $y_{ij}$~($i,j\in\{ 1,2\}$). 
    (b)~Illustration of the derivation of $Y_\mathrm{eff}$. Definitions of $Y_k$~($k\in\{\mathrm{eff},\ \mathrm{L},\ \mathrm{R},\ \mathrm{idler}\}$) are mentioned in Appendix~\ref{appendix_modulate_inductor}.
    }
    \label{sfig_matrix}
\end{figure}

Here, we provide the derivation transforming a modulated inductor to an effective time-independent inductor.
The voltage $V(t)$ across the modulated inductor $L(t)$ is given by
\begin{equation}
\begin{aligned}
    V(t) & = \frac{V_\mathrm{s} e^{i\omega_\mathrm{s}t} + V_\mathrm{i} e^{i\omega_\mathrm{i}t} + V_\mathrm{p} e^{i\omega_\mathrm{p}t}}{2}+ \mathrm{c.c.}\\& = \frac{d}{dt}[L(t) I(t)] \\
    &=\frac{d}{dt}\Bigg[ \left( L_0 + \frac{\delta L e^{i\omega_\mathrm{p}t} + \delta L^\ast e^{-i\omega_\mathrm{p}t}}{2}\right) \\
    &\ \ \ \ \ \ \,\times \left( \frac{I_\mathrm{s} e^{i\omega_\mathrm{s}t} + I_\mathrm{i} e^{i\omega_\mathrm{i}t} + I_\mathrm{p} e^{i\omega_\mathrm{p}t}}{2} + \mathrm{c.c.}\right)\Bigg],
\end{aligned}\label{eq_s1}
\end{equation}
where the current $I(t)$ and the voltage $V(t)$ both have three frequency components~($\omega_\mathrm{s}$, $\omega_\mathrm{i}$ and $\omega_\mathrm{p}$) with complex amplitudes $I_k$ and $V_k$~($k\in\{\mathrm{s},\mathrm{i},\mathrm{p}\}$), respectively.
Note that we assume there are no other mixing products for simplicity.

Grouping terms with the same $e^{i\omega_\mathrm{s}t}$ and $e^{-i\omega_\mathrm{i}t}$ in Eq.~(\ref{eq_s1}), we obtain
\begin{align}
    V_\mathrm{s} &= i\omega_\mathrm{s} L_0 I_\mathrm{s} + i\omega_\mathrm{s}\frac{\delta L}{2}I_\mathrm{i}^\ast,\\
    V_\mathrm{i}^\ast &= -i\omega_\mathrm{i}\frac{\delta L^\ast}{2}I_\mathrm{s} -i\omega_\mathrm{i} L_0 I_\mathrm{i},
\end{align}
which yields the following impedance matrix in the basis of ($V_\mathrm{s},\ V_\mathrm{i}^\ast$) and ($I_\mathrm{s},\ I_\mathrm{i}^\ast$):
\begin{equation}
    \mathbf{Z} = \begin{pmatrix}
    z_{11} & z_{12}\\
    z_{21} & z_{22}
    \end{pmatrix} = \begin{pmatrix}
    i\omega_\mathrm{s} L_0 & i\omega_\mathrm{s} \delta L/2\\
    -i\omega_\mathrm{i} \delta L^\ast/2 & -i\omega_\mathrm{i} L_0
    \end{pmatrix}.\label{eq_Z_matrix}
\end{equation}
We transform the impedance matrix into an admittance matrix $\mathbf{Y}$ and an ABCD matrix as follows:
\begin{align}
    \mathbf{Y} &= \begin{pmatrix}
    z_{22}/|z| & -z_{12}/|z|\\
    -z_{21}/|z| & z_{11}/|z|
    \end{pmatrix} = \begin{pmatrix}
    y_{11} & y_{12}\\
    y_{21} & y_{11}
    \end{pmatrix}, \\
    \begin{pmatrix}
    A & B\\
    C & D
    \end{pmatrix} &= \begin{pmatrix}
    1 & 0\\
    y_{11} & 1
    \end{pmatrix}\!\!\begin{pmatrix}
    0 & -1/y_{21}\\
    y_{12} & 0
    \end{pmatrix}\!\!\begin{pmatrix}
    1 & 0\\
    y_{22} & 1
    \end{pmatrix},\label{eq_abcd}
\end{align}
where \begin{align}
    |z| &= z_{11}z_{22} - z_{12}z_{21} = \omega_\mathrm{s} \omega_\mathrm{i} L_0 L_0^\prime,\\
    y_{11} &= \frac{1}{i \omega_\mathrm{s} L_0^\prime},\\
    y_{22} &= \frac{1}{-i \omega_\mathrm{i} L_0^\prime},\\
    y_{12} & = \frac{\delta L}{L_0}\frac{1}{i \omega_\mathrm{s} L_0^\prime},\\
    y_{21} & = \frac{\delta L^\ast}{L_0}\frac{1}{i \omega_\mathrm{i} L_0^\prime}.
\end{align}

The three matrices on the right-hand side of Eq.~(\ref{eq_abcd}) correspond to a parallel inductor $L_0^\prime$ at $\omega_\mathrm{s}$, an amplification inverter matrix~($\mathbf{J}_\mathrm{PA}$) and another parallel inductor $L_0^\prime$ at $-\omega_\mathrm{i}$, as shown in Figs.~\ref{fig0}(b) and~\ref{sfig_matrix}.
To be consistent with Ref.~\citenum{naaman2022synthesis}, we rewrite $\mathbf{J}_\mathrm{PA}$ as 
\begin{equation}
    \mathbf{J}_\mathrm{PA} = \begin{pmatrix}
    0 & -1/y_{21}\\
    y_{12} & 0
    \end{pmatrix} = \begin{pmatrix}
    0 & i/J_\mathrm{s}^\ast\\
    -i J_\mathrm{i} & 0
    \end{pmatrix},
\end{equation}
where 
\begin{equation}
    J_k = \frac{\sqrt{\alpha}\, e^{i\varphi_\mathrm{p}}}{\omega_k L_0^\prime} \,\, (k \in \{\mathrm{s}, \mathrm{p}\}),
\end{equation}
and $\varphi_\mathrm{p}$ is the phase of $\delta L$.

We aim to simplify the complex circuit into a single element with admittance $Y_\mathrm{eff}$. 
Initially, we compute the admittance $Y_\mathrm{idler}$ seen from $L_0^\prime$ to the idler matching network as defined in Figs.~\ref{fig0}(b) and~\ref{sfig_matrix}(b), including a capacitor, $Z_\mathrm{KI,\lambda/4}$ line, $Z_\mathrm{\lambda/2}$ line, $Z_\mathrm{\lambda/4}$ line and $Z_0$ source, subsequently in our case.
Next, we calculate the admittance on both sides of $\mathbf{J}_\mathrm{PA}$ using $Y_m = I_m/V_m$~($m\in\{\mathrm{L},\mathrm{R}\}$), where $I_m$ and $V_m$ are specified in Fig.~\ref{sfig_matrix}(b).
According to the ABCD matrix formalism, $V_\mathrm{L} = B  I_\mathrm{R} = i I_\mathrm{R} / J_\mathrm{s}$ and $I_\mathrm{L} = C  V_\mathrm{R} = -i J_\mathrm{i} V_\mathrm{R}$, leading to 
\begin{equation} Y_\mathrm{L} Y_\mathrm{R} = -J_\mathrm{s}^\ast J_\mathrm{i}. \end{equation} 
We then transform the admittance from right to left as shown in Fig.~\ref{sfig_matrix}(b): 
\begin{align} Y_\mathrm{R}(-\omega_\mathrm{i}) &= Y_\mathrm{idler}(-\omega_\mathrm{i}) + \frac{1}{-i \omega_\mathrm{i} L_0^\prime}, \\ Y_\mathrm{L}(\omega_\mathrm{s}) &= \frac{\alpha}{\omega_\mathrm{s} \omega_\mathrm{i} {L_0^\prime}^2 Y_\mathrm{R}(-\omega_\mathrm{i})}, \\ Y_\mathrm{eff}(\omega_\mathrm{s}) &= Y_\mathrm{L}(-\omega_\mathrm{i}) + \frac{1}{i \omega_\mathrm{s} L_0^\prime} \nonumber, \\ &=\frac{1}{i \omega_\mathrm{s} L_0^\prime}\left[1+\frac{\alpha}{i \omega_\mathrm{i} L_0^\prime  Y_\mathrm{idler}^\ast(\omega_\mathrm{i}) - 1}\right]\\
&= \frac{1}{i \omega_\mathrm{s} L_\mathrm{eff}},
\end{align}
where $L_\mathrm{eff}$ is the effective inductance associated with the modulated inductance $L(t)$, permitting complex values.
From the aforementioned observation, we find that the real part of $Y_\mathrm{eff}$ is approximately proportional to $-\alpha$.  
Later in Appendix~\ref{appendix_simulation}, we numerically demonstrate this relationship and offer an intuitive understanding of the achievable bandwidth.

\subsection{Three-stage impedance-transformer design}
\label{appendix_network}
\begin{figure}
    \centering
    \includegraphics[width = 7.5cm]{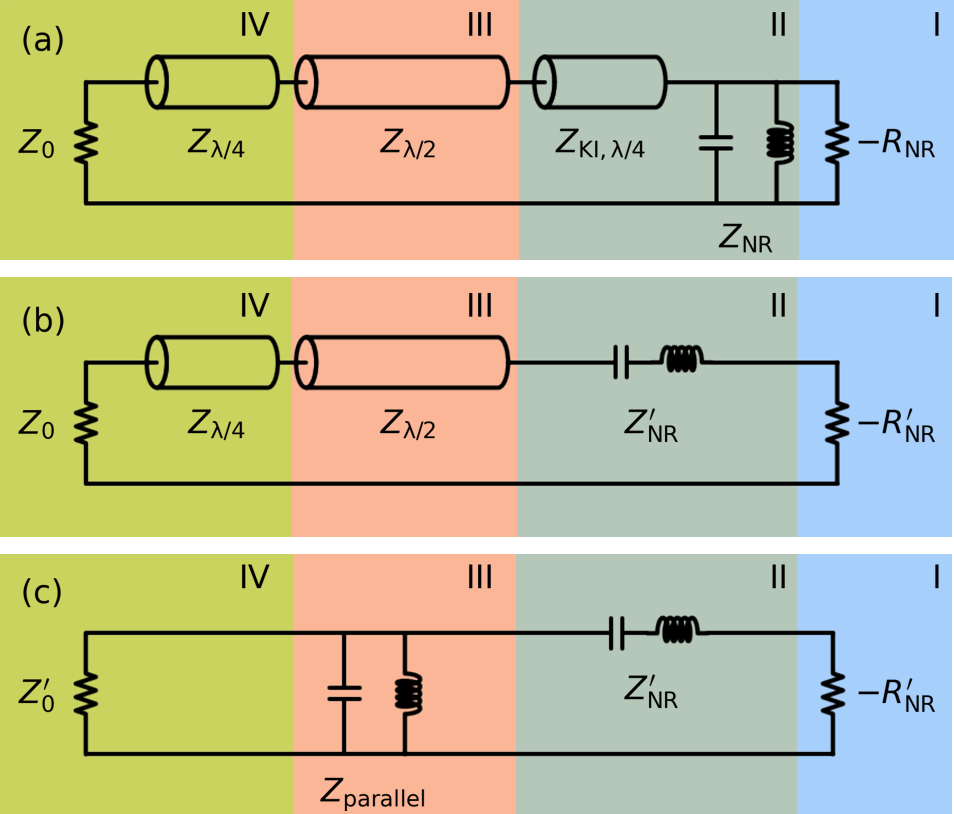}
    \caption{Illustrations of the circuit transformation.
    (a)~Three-stage impedance transformer.    
    The time-dependent inductor under a frequency modulation is linearized into time-independent elements: an inductor in parallel with negative resistance $-R_{\mathrm{NR}}$.
    (b)~Intermediate step transferring the parallel RLC circuit and the $Z_{\mathrm{KI},\lambda/4}$ line into a series RLC circuit in segments~I and~II.
    (c)~Classical lumped-element band-pass circuit. 
    We transfer the $Z_{\lambda/4}$ line and the input source $Z_0 = 50\ \Omega$ into $Z_0^\prime$. 
    Additionally, $Z_{\lambda/2}$ is transferred into a parallel LC circuit~($Z_\mathrm{parallel}$).
    The characteristic impedance values in each segment (I--IV) relate to a given prototype coefficient set $g_i$~($i=0,\ldots,3$) [see Eqs.~(\ref{eqs.g_i_1})--(\ref{eqs.g_i})].
    }
    \label{sfig_network}
\end{figure}
Signal amplification is achieved after transmission into a reflection-type device with negative resistance~\cite{getsinger1963prototypes}.
Appendix~\ref{appendix_modulate_inductor} demonstrated that a simplified complex inductance could represent the modulated inductance, possibly offering negative resistance under a proper circuit design and pump condition. 
Here, we will examine the parametric amplifiers in synthesized impedance-matching networks of our circuit, as depicted in Fig.~\ref{sfig_network}, following Ref.~\citenum{naaman2022synthesis}.

Beginning with the three-stage impedance-transformer amplification circuit in Fig.~\ref{fig0}(a), containing transmission lines of specific lengths corresponding to a wavelength $\lambda$, we substitute $L(t)$ with $L_\mathrm{eff}$ and replace $L_\mathrm{eff}$ again with an inductance in parallel with a negative resistance of value $-R_\mathrm{NR}$ with $R_\mathrm{NR} > 0$, depicted in Fig.~\ref{sfig_network}(a).
Our objective is to derive an equivalent lumped-element band-pass network, including a source resistance of $Z_0^\prime$, two resonators~(a series and a parallel LC circuit), and a negative resistance, $-R_\mathrm{NR}^\prime$, as shown in Fig.~\ref{sfig_network}(c).

We start by examining the impedance seen through a transmission line toward a load $Z_\mathrm{L}$~\cite{pozar2021microwave}:
\begin{equation}
    Z_{\mathrm{tl}\rightarrow \mathrm{L}}(\omega)=Z_{\mathrm{tl}}\frac{Z_\mathrm{L}+iZ_{\mathrm{tl}}\tan\!{\left(\frac{\omega}{v_p}l \right)}}{Z_{\mathrm{tl}}+iZ_\mathrm{L}\tan\!{\left(\frac{\omega}{v_p} l\right)}},\label{eq_see_through}
\end{equation}
where the transmission line has impedance $Z_{\mathrm{tl}}$ and length $l$, and $v_p$ denotes the phase velocity.
When a frequency $\omega$ is close to their fundamental-mode frequency $\omega_0 = 2\pi v_p / \lambda$, and $l$ is equal to $\lambda/2$ or $\lambda/4$, we employ the following approximation:
\begin{equation}
    \tan\!{\left(\frac{\omega}{v_p} l\right)} =
    \begin{cases} 
    \ \ \ \dfrac{\pi \delta \omega}{\omega_0} = \pi\varepsilon & \text{for } l = \frac{\lambda}{2} 
    \\[1em] - \dfrac{2\pi\omega_0 }{\delta\omega} = -\dfrac{2\pi}{\varepsilon} & \text{for } l = \frac{\lambda}{4},
    \end{cases}
\end{equation}
where $\delta\omega = \omega-\omega_0$ and a normalized dimensionless parameter $\varepsilon = \delta \omega/{\omega_0}$.

Next, taking the limit of $\delta\omega\rightarrow 0$ in Eq.~(\ref{eq_see_through}) for a quarter-wavelength transmission line, we obtain $Z_{\lambda/4\rightarrow \mathrm{L}}$ = $Z_{\mathrm{tl}}^2/{Z_\mathrm{L}}$, which is known as the impedance inverter~\cite{pozar2021microwave}.
Similarly, we apply the same rule for segments~I and~II in Fig.~7(a), where the impedance seen through the $Z_{\mathrm{KI},\lambda/4}$ line toward the right-side parallel RLC circuit with an impedance $Z_\mathrm{NR}$ and $R_\mathrm{NR}$ is equivalent to a series RLC circuit with $R_{\mathrm{NR}}^\prime = Z_{\mathrm{KI},\lambda/4}^2 / R_{\mathrm{NR}} $ and a new series-LC impedance $Z_{\mathrm{NR}}^\prime = Z_{\mathrm{KI},\lambda/4}^2 / Z_{\mathrm{NR}}$.

\begin{figure*}[tb]
    \includegraphics[width = 13.78cm]{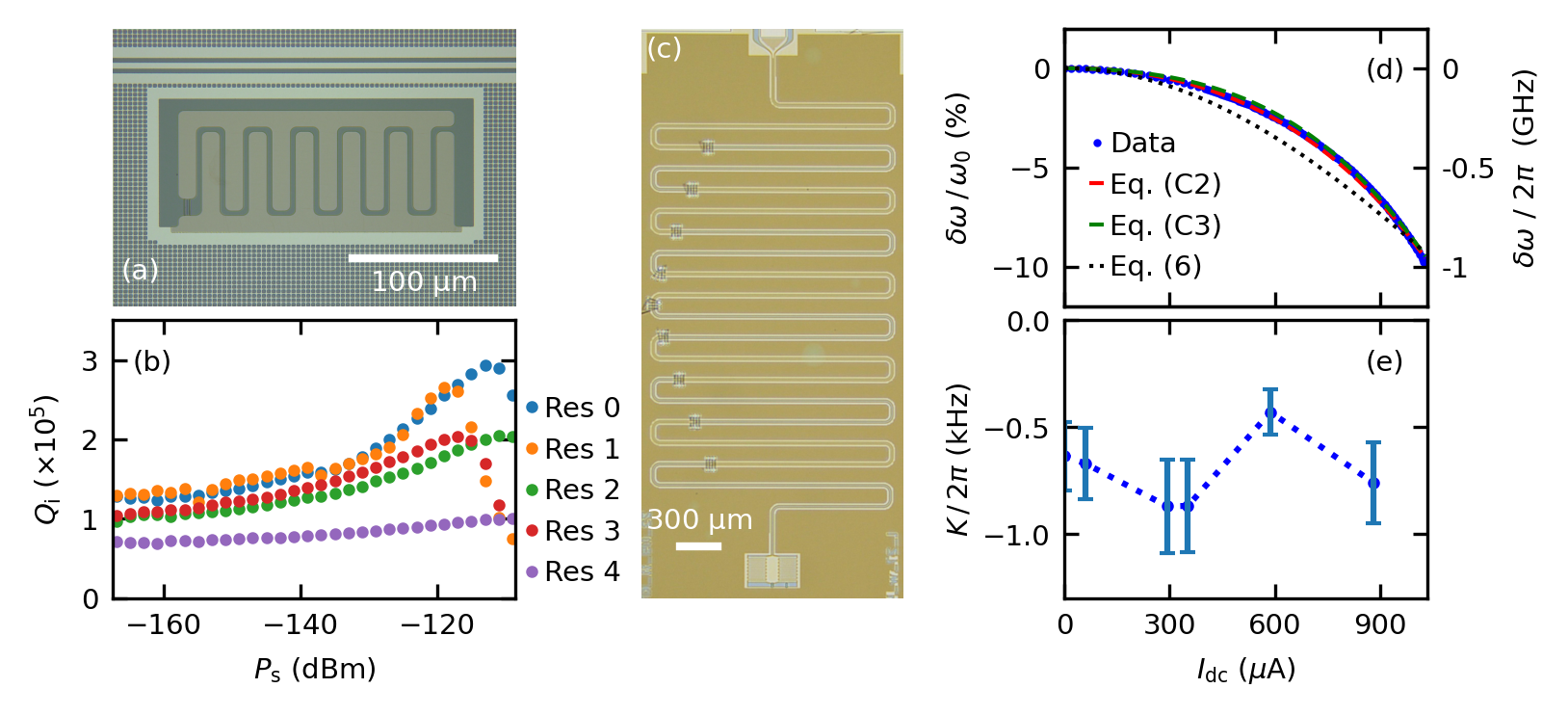}
    \caption{(a) Optical image of a lumped-element LC resonator with a nanowire to calibrate the internal quality factor $Q_\mathrm{i}$. (b)~$Q_\mathrm{i}$ as a function of the signal power $P_\mathrm{s}$ for five different resonators Res 0--4. At the single-photon regime, we observe $Q_\mathrm{i}$ around $10^5$ giving evidence that extra fabrication steps maintain the same quality of thin NbTiN films. (c)~Single-pole kinetic-inductance parametric amplifier. Ten $\lambda/4$ CPWs reduce the coupling between the nonlinear resonator to the 50-$\Omega$ port.
    Unlike the device in~(a), aluminum does not cover the ground plane, the transmission lines or the capacitor in~(c).
    (d)~Resonance-frequency shift $\delta\omega$ vs dc bias current $I_\mathrm{dc}$ of the device shown in~(c).
    Fits from Eqs.~(\ref{eq_fit_reg}),~(\ref{eq_fit_com}) and~(\ref{eq.Lk_approx}) are plotted together. 
    (e)~Kerr coefficient $K$ vs $I_\mathrm{dc}$.}
    \label{sfig_NbTiN}
\end{figure*}

Third, our goal is to match the $Z_{\lambda/2}$ line [Fig.~\ref{sfig_network}(b)] to an equivalent parallel LC circuit with impedance $Z_{\mathrm{parallel}}$~[Fig.~\ref{sfig_network}(c)].
In particular, we calculate the impedance seen through the $Z_{\lambda/2}$ line toward the $Z_{\lambda/4}$ line and $Z_0$.
The impedance seen through the $Z_{\lambda/4}$ line toward the left-side $Z_0$ in segment IV is given by
\begin{equation}
\begin{aligned}
    Z_{\mathrm{IV}} = Z_{\lambda/4\rightarrow Z_0} &\approx \frac{\frac{Z_{\lambda/4}^2}{Z_0} + i Z_{\lambda/4} \frac{\pi \varepsilon}{2}}{1 + i \frac{Z_{\lambda/4}}{Z_0} \frac{\pi \varepsilon}{2}} \\ &= \frac{Z_0^\prime + i \frac{\pi \varepsilon}{2} Z_{\lambda/4}}{1 + i \frac{\pi \varepsilon}{2} \frac{Z_{\lambda/4}}{Z_0}},
\end{aligned}
\end{equation}
where we realize the source $Z_0^\prime = Z_{\lambda/4}^2 / Z_0$ in Fig.~\ref{sfig_network}(c).
Next, the impedance seen from the $Z_{\lambda/2}$ line in segment~III toward segment~IV is 
\begin{equation}
\begin{aligned}
    Z_{\lambda/2\rightarrow \mathrm{IV}} &\approx Z_{\lambda/2} \frac{Z_{\mathrm{IV}} + i \pi \varepsilon Z_{\lambda/2} }{Z_{\lambda/2} + i \pi \varepsilon Z_{\mathrm{IV}}} \\
    &\approx Z_{\lambda/2} \frac{Z_0^\prime + j \frac{\pi \varepsilon}{2} \left(Z_{\lambda/4} + 2 Z_{\lambda/2}\right)}{Z_{\lambda/2} + j \frac{\pi \varepsilon}{2} \left(\frac{Z_{\lambda/4} Z_{\lambda/2}}{Z_0} + 2 Z_0^\prime\right)}.
\end{aligned}
\end{equation}

On the other hand, the impedance of the parallel RLC~[in segments~III and~IV in Fig.~\ref{sfig_network}(c)] is approximated as
\begin{equation}
\begin{aligned}
    Z_{\mathrm{RLC}}^{-1} &= \frac{1}{Z_0^\prime} - \frac{1}{i Z_{\mathrm{parallel}}}
    \left\{1 +\varepsilon - \frac{1}{1+\varepsilon} \right\}
    \\ &\approx \frac{1}{Z_0^\prime} + i 
    \frac{2\varepsilon}{Z_{\mathrm{parallel}}}.
\end{aligned}
\end{equation}
In the limit of small $\delta\omega$, $Z_{\lambda/2\rightarrow \mathrm{IV}}$ equals $Z_{\mathrm{RLC}}$. 
By equating their imaginary components, we derive
\begin{equation}
    Z_{\lambda/2}^2 + Z_{\lambda/2} \left(\frac{Z_{\lambda/4}}{2} - \frac{Z_{\lambda/4} Z_0^\prime}{2 Z_0} + \frac{{Z_0^\prime}^2}{\pi} \frac{2}{Z_\mathrm{parallel}}\right) - {Z_0^\prime}^2 = 0. \label{eq_solve_lambda}
\end{equation}

To determine $Z_{\mathrm{parallel}}$, $Z_{\lambda/4}$ and $Z_{\lambda/2}$, it is essential to select a prototype coefficient set $\{g_i\}$~($i=0,\ldots,3$) for a particular gain and ripples~\cite{naaman2022synthesis}.
$g_0$--$g_3$ correspond to the negative resistance~($-R_{\mathrm{NR}}^\prime$), series LC~(with an impedance of $Z_{\mathrm{NR}}^\prime$), parallel LC~(with an impedance of $Z_{\mathrm{parallel}}$), and the load resistance~($Z_0^\prime$), respectively.
The impedance set of segments I--IV is constructed according to a reference impedance $Z_\mathrm{ref}$~\cite{getsinger1963prototypes}:
\begin{align}
        Z_{\mathrm{ref}} &= g_0R_{\mathrm{NR}}^\prime =\ R_{\mathrm{NR}}^\prime,\label{eqs.g_i_1}\\
        Z_{\mathrm{NR}}^\prime &= \frac{g_1}{\varepsilon}Z_{{\mathrm{ref}}}\label{eq.Z_nr},\\
        \frac{1}{Z_\mathrm{parallel}} &= \frac{g_2}{\varepsilon}\frac{1}{Z_\mathrm{ref}},\\
        Z_0^\prime &= g_3Z_{\mathrm{ref}} = \frac{Z_{\lambda/4}^2}{Z_0}.\label{eqs.g_i}
\end{align}
Utilizing the coefficients from Ref.~\citenum{getsinger1963prototypes}, namely:
\begin{equation}
    \{g_0, g_1, g_2, g_3\} = \{1, 0.408, 0.234, 1.106\},
\end{equation}
and setting $\varepsilon = \delta \omega/{\omega_0} = 500/8000$, $Z_{\mathrm{NR}} = 60\ \Omega$, and $Z_{\mathrm{KI},\lambda/4} = 180\ \Omega$, we compute $Z_{\mathrm{ref}} = 82.7\ \Omega$, $Z_{\lambda/4} = 67.6\ \Omega$, and $Z_{\mathrm{parallel}} = 220.9\ \Omega$. 
Thereafter, solving Eq.~(\ref{eq_solve_lambda}) yields $Z_{\lambda/2} = 33.9\ \Omega$. 
Note that these parameters might not be ideal due to potential inaccuracies from the $\tan\! \left(\omega l / v_p \right)$ approximation.

\section{NbTiN properties}
\label{NbTiN_properties}

Our fabrication process includes ion-milling to create the galvanic contacts of Al with NbTiN and wet etching of Al to expose NbTiN, possibly introducing losses.
To evaluate these losses, we designed five similar lumped-element resonators with resonances near 8.5~GHz.
Each device incorporates a nanowire inductor bridging two NbTiN pads, as illustrated in Fig.~\ref{sfig_NbTiN}(a).
The transmission lines and the ground plane are coated with thick Al, whereas the Al layer that covers the resonator pads is selectively eliminated.
We measure these resonators in a millikelvin environment.
The internal quality factor $Q_\mathrm{i}$ vs the signal power $P_\mathrm{s}$ is shown in Fig.~\ref{sfig_NbTiN}(b). 
When $P_\mathrm{s}$ is low, $Q_\mathrm{i}$ is approximately $10^5$.
$Q_\mathrm{i}$ is enhanced as $P_\mathrm{s}$ increases due to the saturation of two-level systems responsible for dielectric loss~\cite{pappas2011two}. 
This confirms the preservation of good quality in NbTiN films after the ion-milling and the wet-etching processes. 
Notably, flux-trapping holes all over the ground plane~[Fig.~\ref{fig2}(d)] are critical for enhancing $Q_\mathrm{i}$; without these holes, $Q_\mathrm{i}$ fell below $10^4$.

Current sensitivity $I_\ast$ is extracted from the resonance-frequency response while varying the injection of the dc current $I_\mathrm{dc}$ in a different device as shown in Fig.~\ref{sfig_NbTiN}(c).
The design incorporates $N=10$ sections of stepped-impedance $\lambda/4$ transmission-line filters with alternating high~($Z_\mathrm{H}$) and low~($Z_\mathrm{L}$) impedance sections, connected in series to an LC resonator comprised of a nanowire NbTiN and an interdigitated capacitor.
The external quality factor $Q_\mathrm{e}$ is derived to be~\cite{parker2022degenerate}
\begin{equation}
    Q_\mathrm{e} = \left(\frac{Z_\mathrm{H}}{Z_\mathrm{L}}\right)^{\! 2N} \frac{ \pi Z_0}{4 Z_{\mathrm{NR}}}.
\end{equation}
We choose $N$ = 5, $Z_\mathrm{H}$ = 90~$\Omega$, $Z_\mathrm{L}=35\ \Omega$ and $Z_{\mathrm{NR}} = 60\ \Omega$ expecting $Q_\mathrm{e} = 8269$ and measure $Q_\mathrm{e}= 7500$, in good agreement with the theory.
To enhance the accuracy of resonance-frequency measurement, $Q_\mathrm{e}$ is designed to be close to $Q_\mathrm{i}$.

The resonance frequency shift $\delta\omega$ vs $I_\mathrm{dc}$ is shown in Fig.~\ref{sfig_NbTiN}(d).
However, we realize that the fit to Eq.~(\ref{eq.Lk_approx})~(black dotted line) with only the quadratic term fails to match our measurement results.
Taking into account the quartic term $I_{\ast 4}$~\cite{zmuidzinas2012superconducting}, we write 
\begin{equation}
    L_\mathrm{k}(I_\mathrm{dc}) = L_\mathrm{k0}\left[1 + \left(\frac{I_\mathrm{dc}}{I_{\ast 2}}\right)^{\!2} + \left(\frac{I_\mathrm{dc}}{I_{\ast 4}}\right)^{\!4} \right]. \label{eq_fit_reg}
\end{equation}
From the fit, we obtain $I_{\ast 2}$ = 3.25~mA and $I_{\ast 4} = 1.7\ \mathrm{mA}$, while $I_\mathrm{c} = 1.15\ \mathrm{mA}$ is measured by increasing $I_\mathrm{dc}$ until the superconductivity breaks down. 
This yields $I_\mathrm{c}/I_{\ast 2} = 0.35$, which is close to the theoretical limit of 0.42~\cite{zmuidzinas2012superconducting}.

It is worth mentioning another expression $L_\mathrm{k}(I)$, which discusses the behavior of Cooper pairs under both ac and dc currents:
\begin{equation}
    L_\mathrm{k}(I_\mathrm{dc}) = \frac{L_\mathrm{k0}}{\left[1  - (I_\mathrm{dc}/I_{\ast\ast})^{n}\right]^{\frac{1}{n}}},\label{eq_fit_com}
\end{equation} 
where $n$ = 2.21~\cite{clem2012kinetic}.
From this expression, we obtain $I_{\ast\ast} = 1.65\ \mathrm{mA}$ and $I_\mathrm{c}/I_{\ast\ast} = 0.7$, which are in good agreement with the theory suggesting $I_\mathrm{c} \approx 0.66 I_{\ast\ast} $. Eq.~(\ref{eq_fit_com}) is useful since it has only one fitting parameter $I_{\ast\ast}$. 

We determine the surface impedance per square $L_\mathrm{sq} = L_\mathrm{k0}  (w / l)$ by comparing the measured $\lambda/2$ NbTiN CPW resonance to the simulations performed in COMSOL, yielding $L_\mathrm{sq}$ = 6.7~pH/square for the 20-nm-thick film. 
This result is consistent with the measurements of $L_\mathrm{k}(I_\mathrm{dc})$, assuming a fixed geometric inductance $L_\mathrm{g} \approx$ 200~pH according to COMSOL simulations.
\section{Kinetic-inductance nonlinearity}
\label{appendix_KI}
In this Appendix, we will express the inductance modulation amplitude $\delta L$ in terms of the pump current $I_\mathrm{p}$ and $I_\ast$ using the Hamiltonian. 
For simplicity, we begin with Eq.~(\ref{eq.Lk_approx}) while excluding the quartic term associated with $I_{\ast 4}$.
For a parametric oscillator comprised of a parallel inductor and a capacitor, the Lagrangian is given by
\begin{equation}
\begin{aligned}
    \mathcal{L} & =\frac{1}{2}L_\mathrm{k}(I) {\dot{q}}^2-\frac{q^2}{2C} \\
    & =\frac{1}{2}L_\mathrm{k0}{\dot{q}}^2\left(1+\left(\frac{I_\mathrm{dc}}{I_\ast}\right)^{\!2}+\frac{2 I_\mathrm{dc} \dot{q}}{I_\ast^2}+\left(\frac{\dot{q}}{I_\ast}\right)^{\!2}\ \right)-\frac{q^2}{2C},
\end{aligned}
\end{equation}
where $I = I_\mathrm{dc} + I_\mathrm{ac}$ and $I_\mathrm{ac} = \dot{q}$ represents the ac current across the inductor and $q$ is the charge on the capacitors.
To simplify the notation, we introduce $L_\mathrm{I}=L_\mathrm{k0}(1+i_\mathrm{dc}^2)$, where $i_\mathrm{dc} = I_\mathrm{dc}/{I_\ast}$.

The conjugate momentum of $q$ is 
\begin{equation}
\begin{aligned}
    \Phi &=\ \frac{\partial\mathcal{L}}{\partial\dot{q}} \\
     &= \frac{1}{2}L_\mathrm{I}\left(2\dot{q}+\frac{6 i_\mathrm{dc}}{1+i_\mathrm{dc}^2}\frac{{\dot{q}}^2}{I_\ast}+\ \frac{4\dot{q}}{1+i_\mathrm{dc}^2}\left(\frac{\dot{q}}{I_\ast}\right)^{\!2}\right). \label{eq_B2}
\end{aligned}
\end{equation}
To express $\dot{q}$ in $\Phi$, we assume $I_\ast\gg\dot{q}$ in the sense that $I_\ast>I_\mathrm{c}>I_\mathrm{dc}+\dot{q}$ such that the second and the third terms in Eq.~(\ref{eq_B2}) are much smaller than the first term.
We rearrange Eq.~(\ref{eq_B2}) into
\begin{equation}
\begin{aligned}
    \dot{q}&=\frac{\Phi}{L_\mathrm{I}}-\frac{3i_\mathrm{dc}\dot{q}^2}{\left(1+i_\mathrm{dc}^2\right)I_\ast} -\frac{2\dot{q}^3}{\left(1+i_\mathrm{dc}^2\right)I_\ast^2} \\
    &=I_\Phi-\mathcal{A}\dot{q}^2-\mathcal{B}{\dot{q}}^3, \label{eq_dotq}
\end{aligned}
\end{equation}
where $I_\Phi=\frac{\Phi}{L_\mathrm{I}}$, $\mathcal{A}=\frac{3 i_\mathrm{dc}}{\left(1+i_\mathrm{dc}^2\right)I_\ast}$, and 
$\mathcal{B} =\frac{2}{\left(1+i_\mathrm{dc}^2\right)I_\ast^2}$.
We iteratively substitute $\dot{q}$ back into the right-hand side~(RHS) of Eq.~(\ref{eq_dotq}) until RHS contains only powers of $I_\Phi$ up to the quartic order~($I_\Phi^4$) and yield
\begin{equation}
    \begin{split}
        {\dot{q}}^2&\approx I_\Phi^2-2\mathcal{A}I_\Phi^3-2\mathcal{B}I_\Phi^4+5\mathcal{A}^2I_\Phi^4,
        \\
        {\dot{q}}^3&\approx I_\Phi^3-3\mathcal{A}I_\Phi^4,
        \\
        {\dot{q}}^4&\approx I_\Phi^4.
    \end{split}
\end{equation}
The Hamiltonian $\mathcal{H}$ is obtained through Legendre transformation:  
\begin{equation}
\begin{aligned}
    \mathcal{H}&=\Phi\dot{q}-\mathcal{L}\\
    &\approx\frac{1}{2}L_\mathrm{I}\left(I_\Phi^2-\frac{2}{3}\mathcal{A}I_\Phi^3+\mathcal{A}^2I_\Phi^4-\frac{1}{2}\mathcal{B}I_\Phi^4\right)+\frac{q^2}{2C}.
\end{aligned}
\end{equation}
Separating $\mathcal{H}$ to a linear harmonic-oscillator Hamiltonian~$\mathcal{H}_0$ and a perturbed nonlinear Hamiltonian~$\mathcal{H}_1$, we find 
\begin{equation}
    \mathcal{H}_0=\frac{\Phi^2}{2L_\mathrm{I}}+\frac{q^2}{2C},
\end{equation}
and 
\begin{equation}
    \begin{split}
        \mathcal{H}_1&=\frac{1}{2}L_\mathrm{I}\left(-\frac{2}{3}\mathcal{A}I_\Phi^3+\mathcal{A}^2I_\Phi^4-\frac{1}{2}\mathcal{B}I_\Phi^4\right)
        \\
        &=-\frac{i_\mathrm{dc}}{\left(1+i_\mathrm{dc}^2\right) L_\mathrm{I}^2 I_\ast}\Phi^3+\frac{-1+8 i_\mathrm{dc}^2}{2\left(1+i_\mathrm{dc}^2\right)^2  L_\mathrm{I}^3 I_\ast^2}\Phi^4. \label{eq_H1}
    \end{split}
\end{equation}

Subsequently, we treat the strong pump $I_\mathrm{p}(t)$ classically, while the remaining degrees of freedom are treated quantum mechanically in $\Phi$:
\begin{equation}
    \Phi=\Phi_\mathrm{p}+\hat{\Phi} = L_\mathrm{I}  |I_\mathrm{p}|\cos(\omega_\mathrm{p}t+\varphi_\mathrm{p})
    +\hat{\Phi}. \label{eq_phi}
\end{equation}
Here, the quantum operator $\hat{\Phi}$ and $\hat{q}$ are defined by the new linear-oscillator Hamiltonian $\mathcal{H}_0^\prime$ ignoring classical terms:
\begin{align}
    \hat{\Phi}&=\sqrt{\frac{\hbar Z_{\mathrm{NR}}}{2}}\,(\hat{a}^\dagger+\hat{a}), \\
    \hat{q}&=i\sqrt{\frac{\hbar}{2 Z_{\mathrm{NR}}}} \,(\hat{a}^\dagger-\hat{a}), \\
    \mathcal{H}_0^\prime &= \hbar\omega_0\left(\hat{a}^\dagger \hat{a} + \frac{1}{2}\right),
\end{align}
where the characteristic impedance of the nonlinear resonator $Z_{\mathrm{NR}}=\sqrt{{L_\mathrm{I}}/{C}}$, the resonance frequency $\omega_{0}={1}/{\sqrt{L_\mathrm{I}C}}$, and the lowering operator $\hat{a}$ are introduced.
In the Heisenberg picture of $\mathcal{H}_0^\prime$, the operator $\hat{a}$ oscillates in time and $\hat{a} = \hat{a}(0)e^{-i\omega_0t}$.
Substituting Eq.~(\ref{eq_phi}) back into Eq.~(\ref{eq_H1}) and assuming $\omega_\mathrm{p} \approx 2\times \omega_0$, we reformulate $\mathcal{H}_1$ by omitting all fast-rotating and constant terms, and obtain:
\begin{equation}
\begin{aligned}
    \frac{\mathcal{H}_1}{\hbar} = &-\frac{3}{4}\frac{i_\mathrm{dc}}{\left(1+i_\mathrm{dc}^2\right)}\frac{|I_\mathrm{p}| Z_{\mathrm{NR}}}{L_\mathrm{I} I_\ast}\left(a^\dagger a^\dagger e^{-i{(\omega}_\mathrm{p} t+\varphi_\mathrm{p})}+\mathrm{h.c.}\right) \\
    &+\frac{3}{4}\frac{8 i_\mathrm{dc}^2-1}{\left(1+i_\mathrm{dc}^2\right)^2}\frac{\hbar Z_{\mathrm{NR}}^2}{L_\mathrm{I}^3 I_\ast^2}\left(a^\dagger a\right)^2 \\ &+ \frac{3}{2}\frac{8 i_\mathrm{dc}^2-1}{\left(1+i_\mathrm{dc}^2\right)^2}\frac{Z_{\mathrm{NR}}|I_\mathrm{p}|^2}{L_\mathrm{I} I_\ast^2}a^\dagger a.
\end{aligned}
\end{equation}
We define the amplification coefficient $\xi_3$ and Kerr strength $K$ for the cubic and quartic  terms of $a$, respectively, and the pump-induced shift $\delta_p$ as
\begin{equation}
    \xi_3=-\frac{3}{2}\frac{I_\mathrm{dc}|I_\mathrm{p}|}{\left(I_\ast^2+I_\mathrm{dc}^2\right)}\omega_{0} e^{-i\varphi_\mathrm{p}}, \label{eq.xi}
\end{equation}
\begin{equation}
    K=\frac{3}{4}\frac{8 I_\mathrm{dc}^2-I_\ast^2}{\left(I_\ast^2+I_\mathrm{dc}^2\right)^2}\frac{\hbar\omega_{0}^2}{L_\mathrm{I}},
\end{equation}
and 
\begin{equation}
    \delta_\mathrm{p} = \frac{3}{2}\frac{8 I_\mathrm{dc}^2-I_\ast^2}{\left(I_\ast^2+I_\mathrm{dc}^2\right)^2}\omega_{0} |I_\mathrm{p}|^2.
\end{equation}
The full Hamiltonian becomes:
\begin{equation}
    \begin{aligned}
    \mathcal{H}/\hbar &= \left(\omega_0-\frac{K}{2}+\delta_\mathrm{p}\right) \hat{a}^\dagger \hat{a} +\\ &\frac{\xi_3}{2} \hat{a}^\dagger \hat{a}^\dagger + \frac{\xi_3^\ast}{2} \hat{a} \hat{a} + K \hat{a}^\dagger \hat{a}^\dagger \hat{a} \hat{a}.
\end{aligned}
\end{equation}

In Fig.~\ref{sfig_NbTiN}(e), $K$ is determined by sending a strong tone at the resonance of the device shown in Fig.~\ref{sfig_NbTiN}(c), while a weak tone is applied to measure the shift of the resonance frequency.
Because of the error bar, the dependence of $K$ on $I_\mathrm{dc}$ is not distinctly observable.
Given $I_\ast = 3.25$~mA, we anticipate $K/2\pi\approx$ 13~Hz, a value that deviates from our observations.
Nevertheless, $K$ is still 2 to 3 orders of magnitude smaller than a typical JJ-based PA.
A small $K$ is the key to achieving high $P_{1\mathrm{dB}}^{\mathrm{in}}$~\cite{zorin2016josephson,frattini2018optimizing}.

As we did in Appendix~\ref{appendix_modulate_inductor}, we assume that voltage and current contain only the dominant terms of $\omega_\mathrm{s}$, $\omega_\mathrm{i}$, and $\omega_\mathrm{p}$ and obtain the following:
\begin{equation}
    V = \frac{1}{2}\left(V_\mathrm{s} e^{i\omega_\mathrm{s} t} + V_\mathrm{i} e^{i\omega_\mathrm{i} t} + {V}_\mathrm{p} e^{i\omega_\mathrm{p} t} + \mathrm{c.c.}\right),
\end{equation}
and 
\begin{equation}
    \dot{q} = \frac{1}{2}\left(I_\mathrm{s} e^{i\omega_\mathrm{s} t} + I_\mathrm{i} e^{i\omega_\mathrm{i} t} + {I}_\mathrm{p}e^{i\omega_\mathrm{p} t} + \mathrm{c.c.}\right).
\end{equation}
Note that voltage $V$ relates to $\Phi$~(and $\dot{q}$) as~\cite{devoret1995quantum}
\begin{equation}
    \Phi(t) = \int_{-\infty}^t V(t^\prime)dt^\prime.
\end{equation}
Under the assumption of $K\ll\xi_3$, from Eq.~(\ref{eq_B2}) we obtain
\begin{align}
    V &= L_\mathrm{I}\left(\ddot{q}+\frac{6 i_\mathrm{dc}}{1+i_\mathrm{dc}^2}\frac{\dot{q}\ddot{q}}{I_\ast}+\ \frac{6}{1+i_\mathrm{dc}^2}\frac{\dot{q}^2\ddot{q}}{I_\ast^2}\right) \nonumber \\
    & \approx L_\mathrm{I}\left(\ddot{q}  - \frac{2 \xi_3}{\omega_0 I_\mathrm{p}^\ast}\dot{q}\ddot{q}\right)\label{eq_here}.
\end{align}
Substituting
\begin{equation}
    \ddot{q} = i\omega_\mathrm{s} \frac{I_\mathrm{s}}{2} e^{i\omega_\mathrm{s} t} + i\omega_\mathrm{i} \frac{I_\mathrm{i}}{2} e^{i\omega_\mathrm{i} t} + i\omega_\mathrm{p} \frac{{I_\mathrm{p}}}{2} e^{i\omega_\mathrm{p} t} + \mathrm{c.c.}
\end{equation}
\begin{figure*}[tb]
    \centering
    \includegraphics[width = 17.7cm]{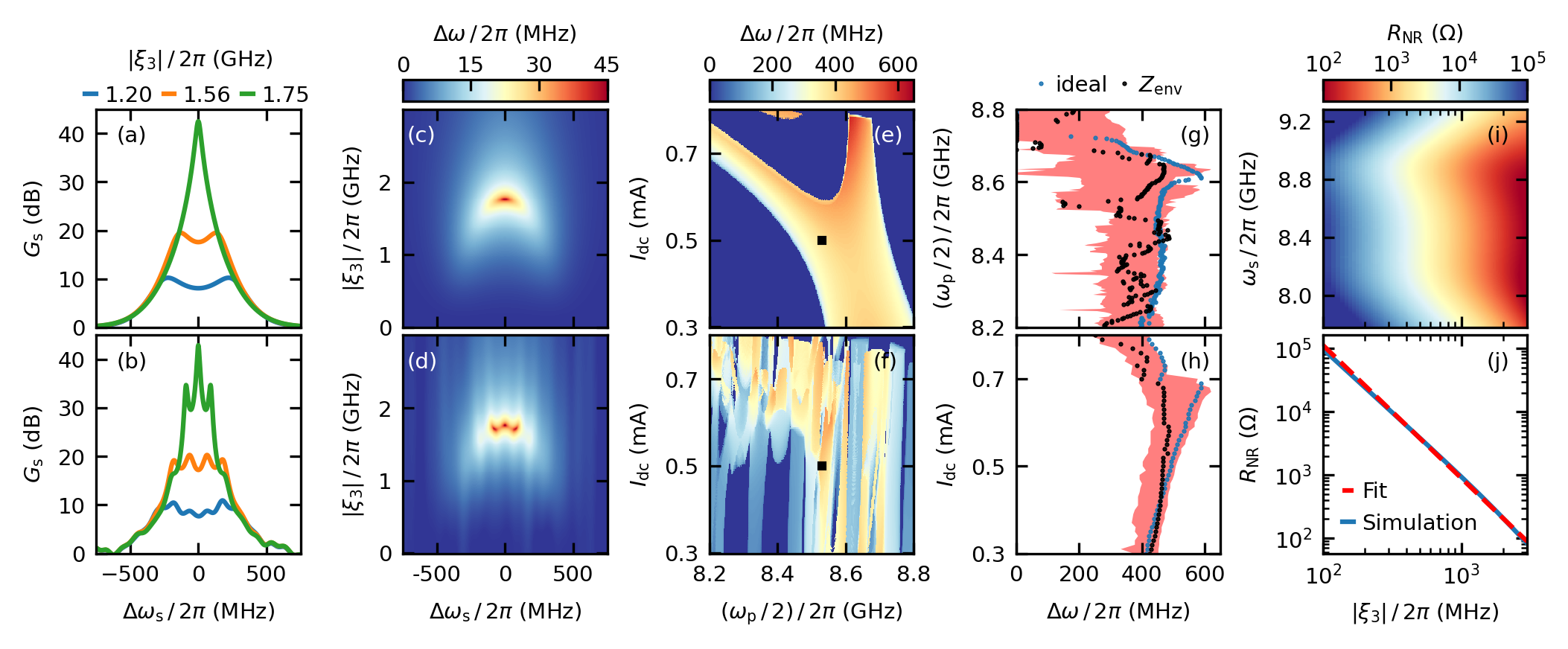}
    \caption{Negative-resistance-circuit simulations. 
    (a),(c),(e)~Simulations with a perfect 50-$\Omega$ environment.
    (b),(d),(f)~Simulations with $Z_\mathrm{env}$ in Eq.~(\ref{eq.z_env}).
    (a)~Signal gain $G_\mathrm{s}$ as a function of the signal detuning $\Delta \omega_\mathrm{s} = \omega_\mathrm{s} -  \omega_\mathrm{p}/2$. At the optimal amplification strength $|\xi_3|/2\pi = $1.56~GHz, we observe two peaks and a bandwidth of 400~MHz at 17-dB gain.
    (b)~$G_\mathrm{s}$ as a function of $\Delta \omega_\mathrm{s}$.
    (c),(d)~$G_\mathrm{s}$ as a function of $\Delta \omega_\mathrm{s}$ and $|\xi_3|$. 
    (e)~Simulation of the bandwidth $\Delta\omega$ at 17-dB gain as a function of the bias current $I_\mathrm{dc}$ and pump frequency $\omega_\mathrm{p}$. The bandwidth at each pair of \{$I_\mathrm{dc}$, $\omega_\mathrm{p}$\} is reported only if the gain ripples are less than 5 dB. 
    (f)~Similar to (e) but under non-ideal environment. 
    The black squares in~(e) and~(f) mark the parameter for panels~(a) to~(d).
    (g)~Maximum bandwidth $\Delta \omega$ as a function of the pump frequency $\omega_\mathrm{p}$. The blue and black dots represent the case of the ideal and non-ideal environments, respectively. The red filled area is the simulations with different phase offset $\phi_1$ in Eq.~(\ref{eq.z_env}). 
    (i)~Negative resistance value $R_\mathrm{NR}$ as a function of $|\xi_3|$ and $\omega_\mathrm{s}$.
    (j)~Negative resistance value when $\omega_\mathrm{s}$ is close to $\omega_\mathrm{p}$. A power-law relation between $R_\mathrm{NR}$ and $|\xi_3|$ is shown with the red fit.
    }
    \label{sfig_simulation}
\end{figure*}
into Eq.~(\ref{eq_here}) and separating different frequency terms, we obtain 
\begin{equation}
    \begin{pmatrix}
    V_\mathrm{s} \\ V_\mathrm{i}^\ast \\ V_\mathrm{i} \\ V_\mathrm{s}^\ast
\end{pmatrix}
 = \begin{pmatrix}
    \mathbf{Z} & 0 \\  0 & \mathbf{Z}^\ast
\end{pmatrix}
\begin{pmatrix}
    I_\mathrm{s} \\ I_\mathrm{i}^\ast \\  I_\mathrm{s}^\ast \\ I_\mathrm{i}
\end{pmatrix},
\end{equation}
where
\begin{align}
    \mathbf{Z} & = i L_\mathrm{I}\begin{pmatrix}
    \omega_\mathrm{s} & \frac{\omega_\mathrm{s}}{2\omega_0}\xi_3^\ast\\[1em]
    -\frac{\omega_\mathrm{i}}{2\omega_0}\xi_3 & -\omega_\mathrm{i}
    \end{pmatrix}.
\end{align}
Comparing to the Z-matrix of Eq.~(\ref{eq_Z_matrix}), we obtain 
\begin{equation}
\begin{split}
    L_\mathrm{I} &= L_\mathrm{k}(I_\mathrm{dc}), \\
    \delta L &= \frac{3}{2}\frac{I_\mathrm{dc} I_\mathrm{p}^\ast}{I_\ast^2+I_\mathrm{dc}^2} L_\mathrm{I}, \\
    \alpha &= \frac{9}{16}\left(\frac{I_\mathrm{dc} |I_\mathrm{p}|}{I_\ast^2+I_\mathrm{dc}^2}\right)^{\!2} = \frac{1}{4}\frac{|\xi_3|^2}{\omega_0^2}.
\end{split}
\end{equation}

For comparison, a single JJ exhibits~\cite{naaman2022synthesis}
\begin{equation}
    L_\mathrm{J}(I) = \frac{L_\mathrm{J0}}{\sqrt{1-\frac{I^2}{I_\mathrm{J,c}^2}}} \approx L_\mathrm{J0}\left( 1 + \frac{I^2}{2 I_\mathrm{J,c}^2} \right)
\end{equation}
and 
\begin{equation}
     \alpha \approx \frac{1}{4}\left(\frac{I_\mathrm{dc}|I_\mathrm{p}|}{I_\mathrm{J,c}^2+I_\mathrm{dc}^2}\right)^{\!2}.
\end{equation}
Since KI materials typically have $I_\ast$ values 2--3 orders of magnitude higher than $I_\mathrm{J,c}$, they require proportionally higher pump currents to reach the same $\alpha$.
We choose nanowire NbTiN inductors~(20\nobreakdash-nm-thick and 250\nobreakdash-nm-wide) to reduce $I_\ast$ to the order of milliampere, further mitigating potential heating.

\section{Negative-resistance simulation}
\label{appendix_simulation}

Here, we simulate the reflection coefficient $S_{11}$ of our three-stage impedance-transformer circuit with a negative resistance by direct calculations based on Eqs.~(\ref{eq_eff_Y}) and~(\ref{eq_see_through}).
Since $|\xi_3|$, which is proportional to pump current $|I_\mathrm{p}|$, provides a clear physical meaning for the strength of amplification, we plot $|\xi_3|$ vs gain instead of $R_{\mathrm{NR}}$ vs gain.
Note that the simulation parameter set of \{$Z_{\mathrm{KI},\lambda/4}$, $Z_{\lambda/4}$, $Z_{\lambda/2}$\} is chosen to have near 500 MHz in the non-ideal environment in our operation range of $I_\mathrm{dc}$ and $\omega_\mathrm{p}$ but still remaining near the designed set.

An example of the signal-gain profile $G_\mathrm{s}$ as a function of the signal detuning $\Delta\omega_\mathrm{s} = \omega_\mathrm{s} - \omega_\mathrm{p}/2$ in an ideal 50-$\Omega$ environment is shown in Fig.~\ref{sfig_simulation}(a).
As the pump power increases, the two peaks grow with the increase of pump power, representing a two-pole system. 
At the optimal pump point, $|\xi_3|/2\pi = 1.56$~GHz, $G_\mathrm{s}$ overcomes the gain--bandwidth product but reduces to only a single peak when it is over-pumped~($|\xi_3|/2\pi = $1.75~GHz).
In Fig.~\ref{sfig_simulation}(c), we show the arc-like feature of a two-pole system.
In Figs.~\ref{sfig_simulation}(b) and~(d), we simulate the gain using $Z_\mathrm{env}$ in Eq.~(\ref{eq.z_env}).
Since the PA interacting with the non-ideal environment is analogous to a $N$-pole system~($N>2$), we observe four peaks matching the measurement in Fig.~\ref{fig2}(c).

In Figs.~\ref{sfig_simulation}(e) and~(f), we simulate the plot depicted in Fig.~\ref{fig2}(e) by sweeping $\omega_\mathrm{p}$ and $I_\mathrm{dc}$ and report the maximum bandwidth $\Delta\omega$ for each $(\omega_\mathrm{p},I_\mathrm{dc})$ pair at 17-dB gain only if the gain profile shows more than one peak and the gain ripples are smaller than 5~dB. 
In panel~(e), the regions with non-zero bandwidth are largely confined to a Y-shaped area. 
Above the Y-shape, a triangular zone shows no bandwidth despite a two-peak feature, due to excessive ripples of $>$5~dB. 
Conversely, in the condition with impedance mismatching in panel~(f), we observe ripples $<$5~dB in the triangular zone.
Figure~\ref{fig2} illustrates the experimental gain profiles across different $I_\mathrm{dc}$ and $\omega_\mathrm{p}$, highlighting more intricate ripple patterns with uneven amplitudes compared to the simulations.
This discrepancy is probably caused by differing coupling strengths in cable modes.
As a result, we did not eliminate the gain data with ripples  $>$5~dB in Fig.~\ref{fig2} and report the highest bandwidth of 470 MHz.
Note that no additional attenuator, which is effective at suppressing ripples yet ill-suited for reflection-type PAs, was introduced between the PA and circulators. Extra attenuation would undesirably decrease signal amplitude at both the input and output in practical measurement environments. 

In Figs.~\ref{sfig_simulation}(g) and~(h), we present the peak bandwidth for various $I_\mathrm{dc}$ and $\omega_\mathrm{p}$, respectively, with data depicted in blue for ideal environments and black for non-ideal ones. 
The red shaded area represents simulations with identical $Z_\mathrm{env}$ but varying phase offsets $\phi_1\in\{-\pi,\pi\}$ in the primary sinusoidal impedance modulation. 
The emergence of additional modes from standing waves in the coaxial cable typically decreases rather than increases the maximum bandwidth.

In Fig.~\ref{sfig_simulation}(i), we report the absolute value of the negative resistance value, $R_\mathrm{NR}$, as a function of $\omega_\mathrm{s}$ and $|\xi_3|$ in the three-stage impedance transformer design. 
A cross-section at $\omega_\mathrm{s} \approx \omega_\mathrm{p}/2$, depicted in Fig.~\ref{sfig_simulation}(j), reveals that $R_\mathrm{NR}$ adheres to a power law, $R_\mathrm{NR} \propto \xi_3^{b}$, where $b = -2.1$.
From Eq.~(\ref{eq.Z_nr}), we obtain that the bandwidth is proportional to $R_\mathrm{NR}$ as 
\begin{equation}
    \begin{split}
    \varepsilon &= \Delta\omega / \omega_0 = g_1Z_\mathrm{NR}R_\mathrm{NR}^{-1} \\ & \propto g_1Z_\mathrm{NR} |\xi_3|^{2.1} \propto |I_\mathrm{p}|^{2.1} \proptosim \alpha. \label{eq.bandwidth_relation}
    \end{split}
\end{equation}
The equation suggests that bandwidth grows with $|\xi_3|$.
There is a trade-off between the heating due to the pump power and the bandwidth.

\section{Comparison to Roy's model}
\label{Appedix_roy}
\begin{figure}
    \centering
    \includegraphics[width = 8.5cm]{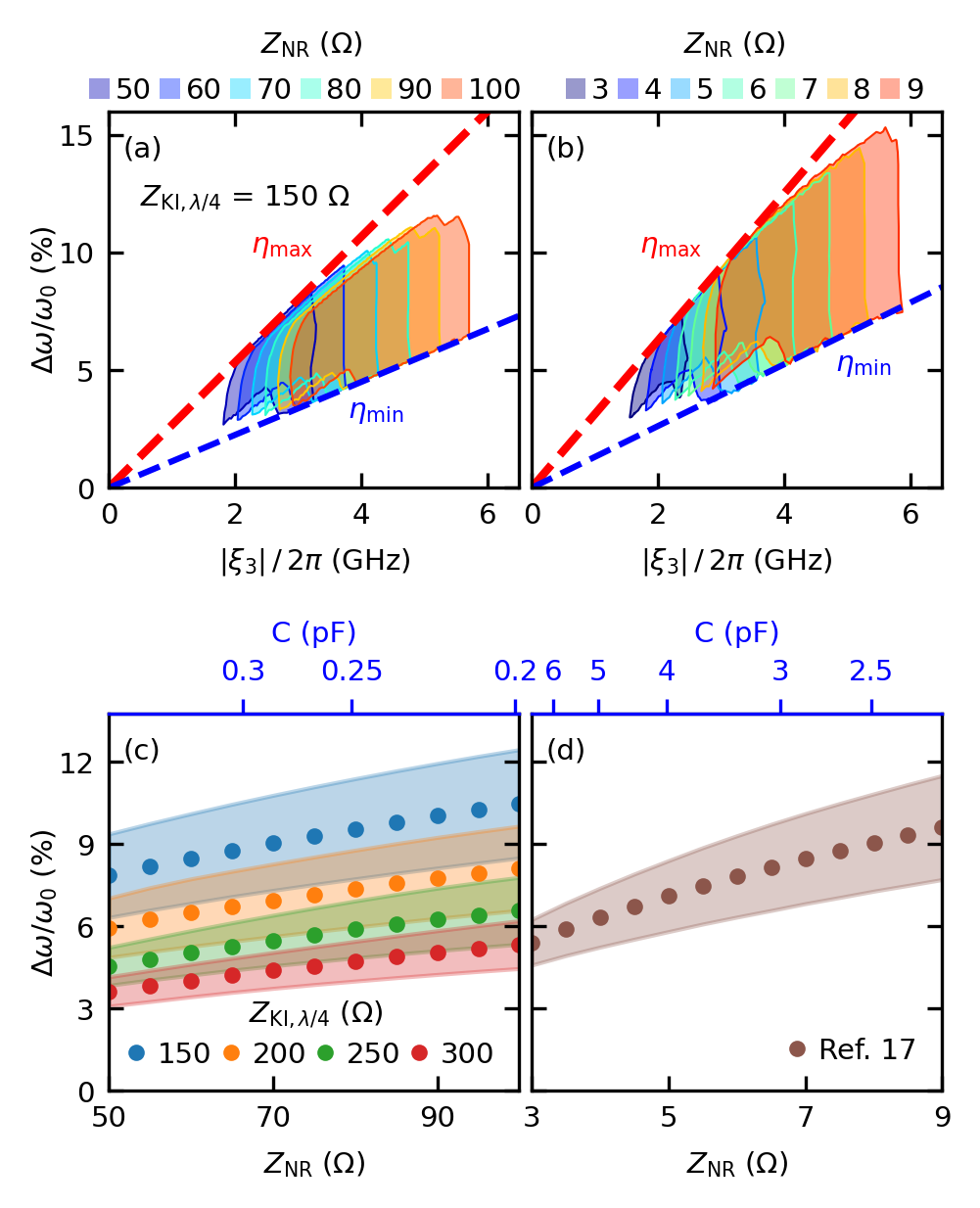}
    \caption{Comparison of our three-stage impedance-transformer and the conventional circuit in Ref.~\citenum{roy2015broadband}.
    (a)~Bandwidth $\Delta\omega$ as a function of $|\xi_3|$ in our circuit, categorized by $Z_\mathrm{NR}$. 
    Two dashed lines represent the maximum~(red) and minimum~(blue) pump efficiency $\eta$ = $\Delta\omega / |\xi_3|$. 
    We observe that maximum $\eta$ has a negative correlation with $Z_{\mathrm{NR}}$.
    (b)~Bandwidth $\Delta\omega$ as a function of $|\xi_3|$ using the conventional circuit~\cite{roy2015broadband}. The same brute-force search process as described in (a) is applied.
    (c)~Average~(dots) and the standard deviation~(shaded area) of $\Delta\omega$ as a function of $Z_{\mathrm{NR}}$ using our circuit. On the upper horizontal axis, we show the required capacitance.
    (d)~Average and the standard deviation of $\Delta\omega$ as a function of $Z_{\mathrm{NR}}$ using the conventional circuit.
    For $\Delta\omega/\omega_0$ = 6\%, our circuit requires only $\sim$400~fF compared to 5~pF in the conventional circuit. }\label{sfig_comparison}
\end{figure}

In Appendix~\ref{appendix_network}, we demonstrated that the three-stage impedance-transformer circuit allows us to use high-impedance parametric oscillators. 
This section systematically compares our circuit to the conventional circuit lacking the $Z_{\mathrm{KI},\lambda/4}$ line~\cite{roy2015broadband}.
We perform a brute-force search over all feasible combinations of the design parameters including $Z_{\lambda/4}$, $Z_{\lambda/2}$, $Z_{\mathrm{NR}}$, and $\omega_\mathrm{p}$. 

\begin{figure*}
    \centering
    \includegraphics[width = 15.4cm]{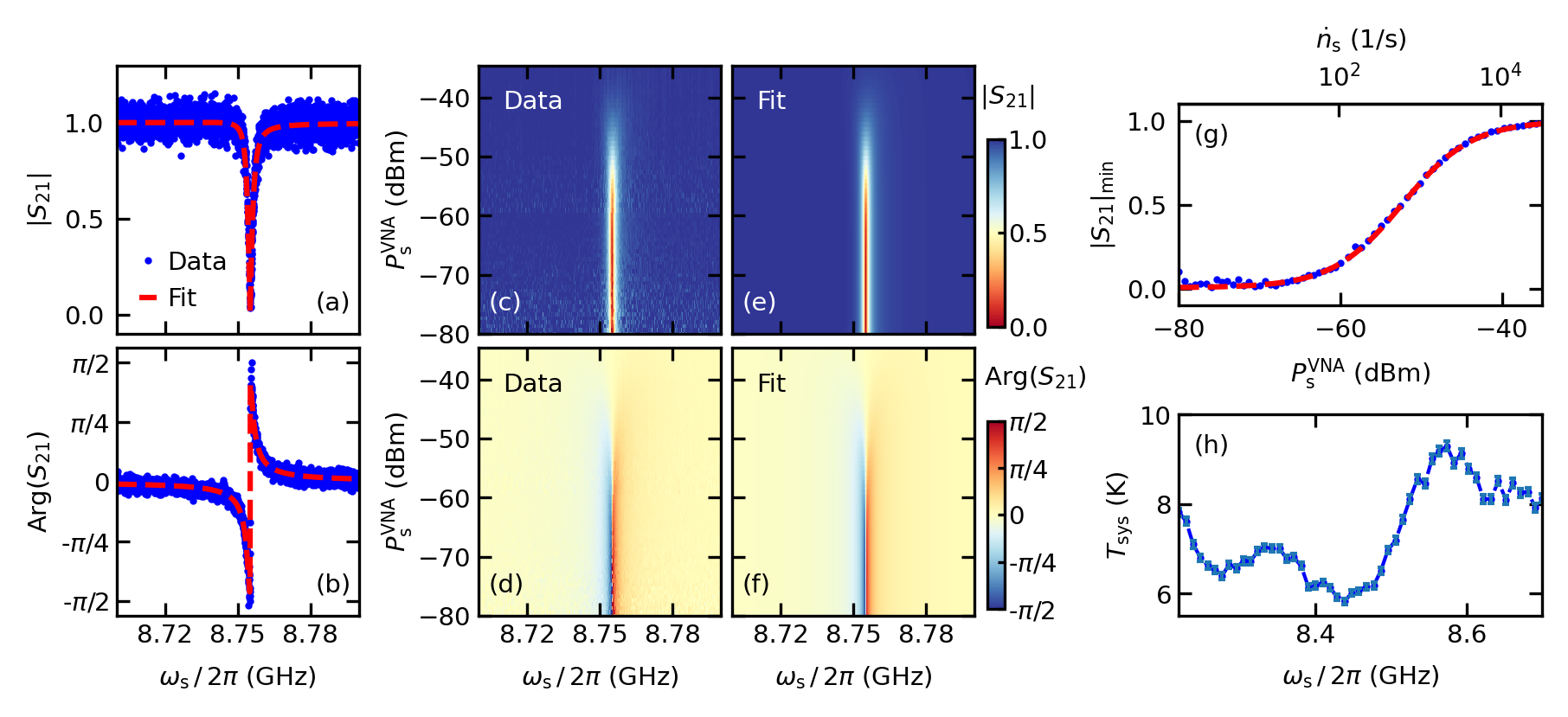}
    \caption{Calibration of the signal power using a transmon qubit coupled to a transmission line. (a) Amplitude and~(b) phase of the transmission coefficient $S_{21}$ as a function of the signal frequency $\omega_\mathrm{s}$. The data is taken at a low signal power from VNA, $P_\mathrm{s}^\mathrm{VNA} = -75$~dBm.
    The red dashed lines are the fit to Eq.~(\ref{eq_qb_fit}).
    (c)~Amplitude and (d)~phase of $S_{21}$ as a function of $\omega_\mathrm{s}$ and $P_\mathrm{s}^\mathrm{VNA}$. 
    (e),~(f)~Fits to (c) and~(d). 
    (g)~$|S_{21}|_\mathrm{min}$ at the resonance frequency as a function of $P_\mathrm{s}^\mathrm{VNA}$. 
    The top axis indicates the corresponding number of signal photons per second, $\dot{n}_\mathrm{s}$. 
    (h) System noise temperature $T_{\mathrm{sys}}$ vs $\omega_\mathrm{s}$.
    }\label{sfig_qb}
\end{figure*}

First, in our circuit, we set fixed values for the nonlinear-oscillator resonance frequency $\omega_0 / 2\pi$ = 8~GHz, $\lambda$ corresponding to $\omega_0$, and $Z_{\mathrm{KI},\lambda/4}$ = 150~$\Omega$.
Then, we sweep multiple parameters across specific ranges: $Z_{\lambda/4} \in [30\,\Omega,\ 100\,\Omega] $, $Z_{\lambda/2} \in [30\,\Omega,\ 100\,\Omega]$, $(\omega_\mathrm{p}/2)/2\pi \in [7.5\,\mathrm{GHz},\ 8.5\,\mathrm{GHz}]$ and $Z_{\mathrm{NR}} \in [50\,\Omega,\ 100\,\Omega]$ .
We simulate the response similar to Fig.~\ref{sfig_simulation}(a) for each parameter set \{$Z_{\lambda/4}$, $Z_{\lambda/2}$, $\omega_\mathrm{p}$, $Z_{\mathrm{NR}}$\} with a perfect impedance matching.
During the increase of $|\xi_3|$ until the gain $G_\mathrm{s}(|\xi_3|)$ surpasses 40~dB, we record the maximum $\Delta\omega$ with a correspondingly optimal $|\xi_3|$ in the condition that $G_\mathrm{s}$ shows more than 17-dB gain with ripples under 5~dB and two local maxima, resembling a two-pole system.
For instance of Fig.~\ref{sfig_simulation}(a), we increase $|\xi_3| /2\pi$ until 2~GHz and observe $G_\mathrm{s}>40$~dB.
The result shows a maximum $\Delta\omega/2\pi \approx 420$~MHz and the corresponding optimal $|\xi_3| /2\pi = 1.5$~GHz.
The enclosed area, categorized by $Z_{\mathrm{NR}}$ using a consistent color as shown in Fig.~\ref{sfig_comparison}(a), indicates the achievable maximum $\Delta\omega$ and optimal $|\xi_3|$.
As $Z_{\mathrm{NR}}$ rises, the corresponding area shifts to the right, implying larger $|\xi_3|$ is needed for larger $Z_{\mathrm{NR}}$.

The pump efficiency $\eta$, defined as $\eta=\Delta\omega/|\xi_3|$, is important in indicating how the pump power transfers to the bandwidth.
Dashed lines indicate the maximum~(red) and minimum~(blue) $\eta$, revealing that low $Z_{\mathrm{NR}}$ has a slightly better pump efficiency.
Furthermore, the figure indicates that for the same $\alpha$~($\propto |\xi_3|^{2.1}$), a nonlinear resonator with reduced $Z_{\mathrm{NR}}$ exhibits an increased $\eta$. 

In Fig.~\ref{sfig_comparison}(b), we extend our analysis to the conventional circuit~\cite{roy2015broadband}, revealing that the allowed $Z_{\mathrm{NR}} \in [3\,\Omega,\ 9\,\Omega]$ is approximately ten times smaller.
On the other hand, the pump efficiency is slightly better than our proposed circuit.
In Fig.~\ref{sfig_comparison}(c), we compare $\Delta\omega / \omega_0$ as a function of $Z_{\mathrm{NR}}$ with various $Z_{\mathrm{KI},\lambda/4}$.
Dots are the average of all possible $\Delta\omega$, and the area shows the standard deviation.
All $Z_{\mathrm{KI},\lambda/4}$ behave similarly that $\Delta\omega$ raises with $Z_{\mathrm{NR}}$.
Since $|\xi_3|$ is proportional to $|I_\mathrm{p}|$~(or the square root of the pump power), if the pump power is constrained by the function generator or $|\xi_3|$ is limited by the material's nature, selecting a smaller $Z_{\mathrm{NR}}$ or a larger $Z_{\mathrm{KI},\lambda/4}$ is a viable strategy for designing PAs in our scheme.
This design principle opposes Eq.~(\ref{eq_bw_alpha}), which asserts that bandwidth should be proportional to $Z_{\mathrm{NR}}$.
This discrepancy arises from the fact that increasing $Z_{\mathrm{NR}}$ under identical impedance-transformer parameters might have larger bandwidth but also larger ripples, which ultimately is not optimal and would be reported as zero bandwidth.

It is worth mentioning that, given the theoretical limitation $I_\mathrm{dc}+|I_\mathrm{p}| < I_\mathrm{c} \lesssim 0.42 I_\ast$~\cite{zmuidzinas2012superconducting}, we can derive the following inequality and calculate the upper limit of $|\xi_3|$:
\begin{equation}
    \frac{|\xi_3|}{\omega_0} = \frac{3}{2}\frac{I_\mathrm{dc}|I_\mathrm{p}|}{\left(I_\ast^2+I_\mathrm{dc}^2\right)} \leq \frac{3}{2}\frac{(I_\mathrm{c}-|I_\mathrm{p}|) |I_\mathrm{p}|}{5.7 I_\mathrm{c}^2+(I_\mathrm{c}-|I_\mathrm{p}|)^2}, \label{eq_xi_limit}
\end{equation}
which yields maximum $|\xi_3|/2\pi \approx 0.6$~GHz.
In our circuit, the simulated maximum $\eta = 0.21$ yields a maximum bandwidth of 120~MHz, which is much smaller than our measurement.
Apparently, we underestimate $|\xi_3|$, and the precise form of $L_\mathrm{k}(I)$ should include higher-order terms and follow the analysis of both effects of dc and ac currents in Ref.~\citenum{clem2012kinetic}, which is beyond the scope of this paper.

\begin{figure}
    \centering
    \includegraphics[width = 6.2cm]{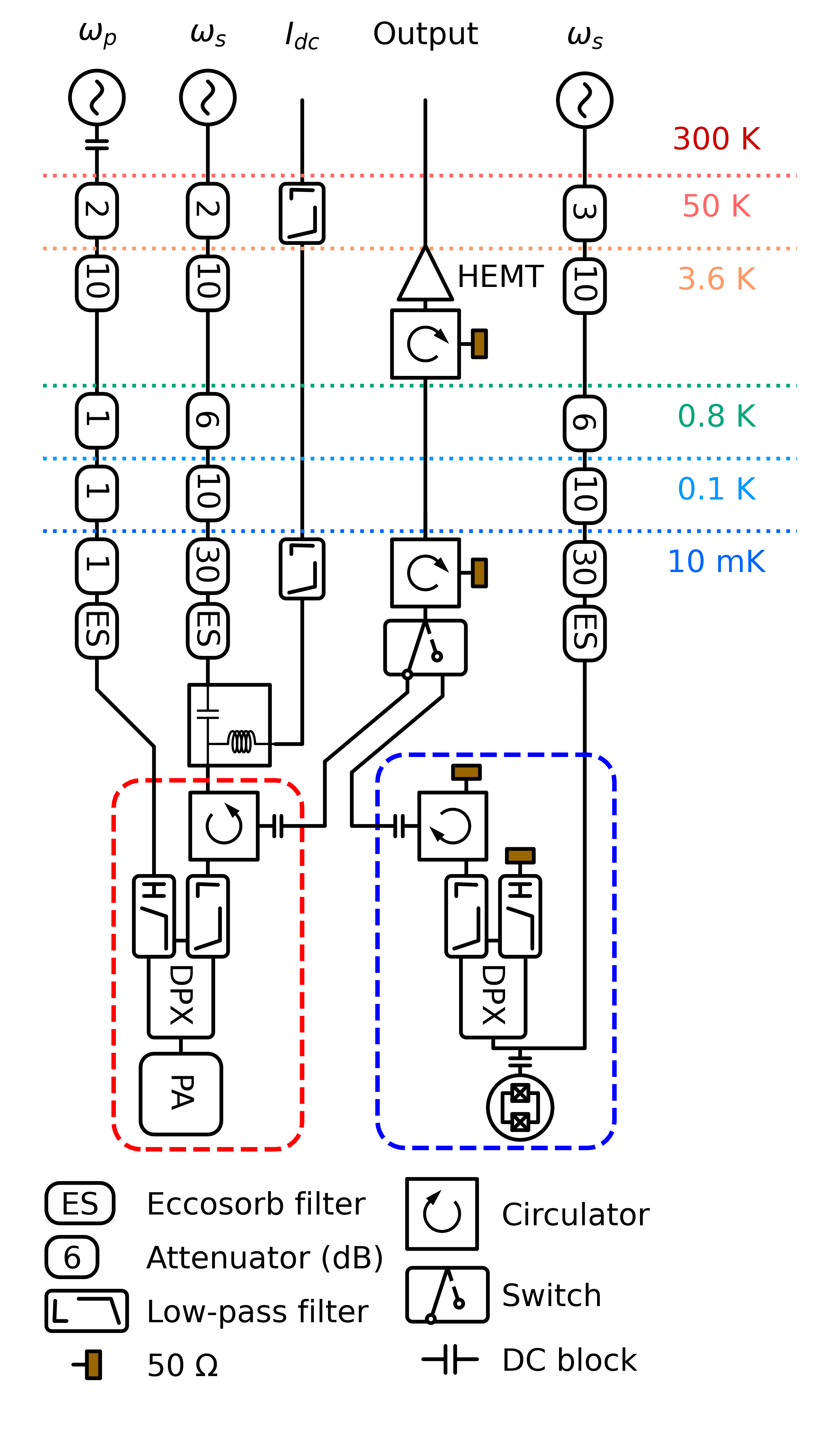}
    \caption{Schematic of the cryogenic microwave-wiring setup. The PA and the transmon qubit share the same output chain via a microwave switch.}\label{sfig_setup}
\end{figure}

\section{System-noise calibration and experimental setup}
\label{appen_sys_noise}

This appendix describes the method for extracting the input-line attenuation $A_\mathrm{in}$ using a qubit.
A qubit coupled to a waveguide with an external coupling rate $\Gamma_{1,\mathrm{E}}$ much larger than its internal decay rate $\Gamma_{1,\mathrm{I}}$ is a useful tool to calibrate the input power.
Under a weak input coherent drive sent through the waveguide, the qubit absorbs a single photon and emits the photon in both directions with a $\pi$ phase shift.
The forward signal interferes destructively with the emission, causing a dip in $|S_{21}|$ near the qubit frequency $\omega_\mathrm{q}$ [Fig.~\ref{sfig_qb}(a)].
On the contrary, with a strong input, the qubit can only absorb one photon before emitting it, and $S_{21}$ approaches unity.
Solving the master equation in the steady state with $\hbar\omega \gg k_\mathrm{B} T$, $S_{21}$ is written as~\cite{astafiev2010resonance,mirhosseini2019cavity}
\begin{equation}
    S_{21}(\omega) = 1 - \frac{\Gamma_{1,\mathrm{E}}}{2\Gamma_2}\frac{1 + i\frac{\Delta_\mathrm{q}}{\Gamma_2}}{1 + \left(\frac{\Delta_\mathrm{q}}{\Gamma_2}\right)^2 + \frac{\Omega_\mathrm{d}^2}{\Gamma_1\Gamma_2}}\label{eq_qb_fit},
\end{equation}
where $\Delta_\mathrm{q} = \omega - \omega_\mathrm{q}$ is the detuning between the input and qubit frequencies, $\Gamma_1 = \Gamma_{1,\mathrm{E}} + \Gamma_{1,\mathrm{I}}$ is the total decay rate, $\Gamma_2 = \Gamma_\phi + \Gamma_1/2$ is the dephasing rate, and $\Gamma_\phi$ is the pure dephasing rate of the qubit.
Here $\Omega_\mathrm{d}$ is the drive strength at the input signal power $P_\mathrm{d}$, and 
\begin{equation}
    \Omega_\mathrm{d} = \sqrt{\frac{2\Gamma_{1,\mathrm{E}}P_\mathrm{d}}{\hbar\omega_\mathrm{q}}}.
\end{equation}
In the case of $\Gamma_1 \approx \Gamma_{1,\mathrm{E}}$ or $\Gamma_{1,\mathrm{E}} \gg \Gamma_{1,\mathrm{I}}$, we fit the measured $S_{21}$ across the power $P_\mathrm{s}^\mathrm{VNA}$ sent from the vector network analyzer at room temperature to extract the input attenuation by $A_\mathrm{in} = P_\mathrm{s}^\mathrm{VNA}/P_\mathrm{d}$.

Measured $S_{21}$ and its fit are shown in Figs.~\ref{sfig_qb}(c)\nobreakdash--(f).
From the fit, we extracted $\Gamma_1/2\pi$ = 3.35$\pm$0.15~MHz,  $\Gamma_\phi/2\pi$ = 1.06$\pm$0.11~kHz, and $\Omega_\mathrm{d}/2\pi$ = 98.6$\pm$9.7~kHz at $P_\mathrm{s}^\mathrm{VNA} = -80\ \mathrm{dBm}$, resulting in $A_\mathrm{in}$ = $-82\pm0.85$~dB.
In Fig.~\ref{sfig_qb}(g), we show the minimum magnitude of $S_{21}$ vs $P_\mathrm{s}^\mathrm{VNA}$ at $\Delta_\mathrm{q}=0$.
On the upper horizontal axis, the photon number per second, $n_\mathrm{s} = \Omega_\mathrm{d}^2/2\Gamma_1$, is shown, where the saturation starts at $n_\mathrm{s}\gg$1 or $P_\mathrm{s}^\mathrm{VNA}\approx$ $-$57~dBm.

In Eq.~(\ref{eq.added_noise}), all variables except $G_\mathrm{sys}^\mathrm{eff} = A_{23} G_{\mathrm{sys}}$ are retrieved using a spectrum analyzer, and $G_\mathrm{sys}^\mathrm{eff}$ is calculated via the equation $P_\mathrm{s}^\mathrm{VNA} A_\mathrm{in} G_\mathrm{s} G_\mathrm{sys}^\mathrm{eff} = P_\mathrm{f}$, where $P_\mathrm{f}$ is the output power measured with the spectrum analyzer. 
As $G_\mathrm{sys}^\mathrm{eff}$ only relates to post-PA components, its determination does not depend on the input-line setup. 
Two input lines are thus utilized for the PA and qubit, sharing an output line, with only one microwave switch for signal routing to minimize the number of switches~(see Fig.~\ref{sfig_setup} for our setup; typically there is another one placed before the PA).
As PA measurements are of the reflection type, the amplified signal undergoes attenuation by a diplexer and a circulator, which contributes to $A_{23}$.
Therefore, this additional loss mechanism must be accounted for in the qubit measurement chain, where a diplexer and a circulator are situated directly after the qubit sample holder. 
This ensures consistent $A_{23}$ across both paths, presuming similar loss effects from these passive components, where the difference between two paths is less than 0.3~dB in the room-temperature transmission measurement. 
However, this difference might increase at cryogenic temperatures, thus a conservative $\pm$0.5-dB error margin is adopted for the noise data in Fig.~\ref{fig4}(d). 
We perform saturation measurements on a tunable qubit at different frequencies by flux biasing from 8.2 to 8.7~GHz to extract $A_\mathrm{in}(\omega)$. 
In Fig.~\ref{sfig_qb}(h), assuming $N_{\mathrm{sys}} \gg N_3^{\mathrm{off}}$, the system noise temperature is expressed as $T_{\mathrm{sys}} = N_4^{\mathrm{off}}\hbar\omega / k_\mathrm{B} G_\mathrm{sys}^\mathrm{eff}$, resulting in $T_{\mathrm{sys}}$ values between 5.9 and 9.2~K.

\section{Different cooldowns and another device}
\label{appen_diff}
\begin{figure}
    \centering
    \includegraphics[width = 8.2cm]{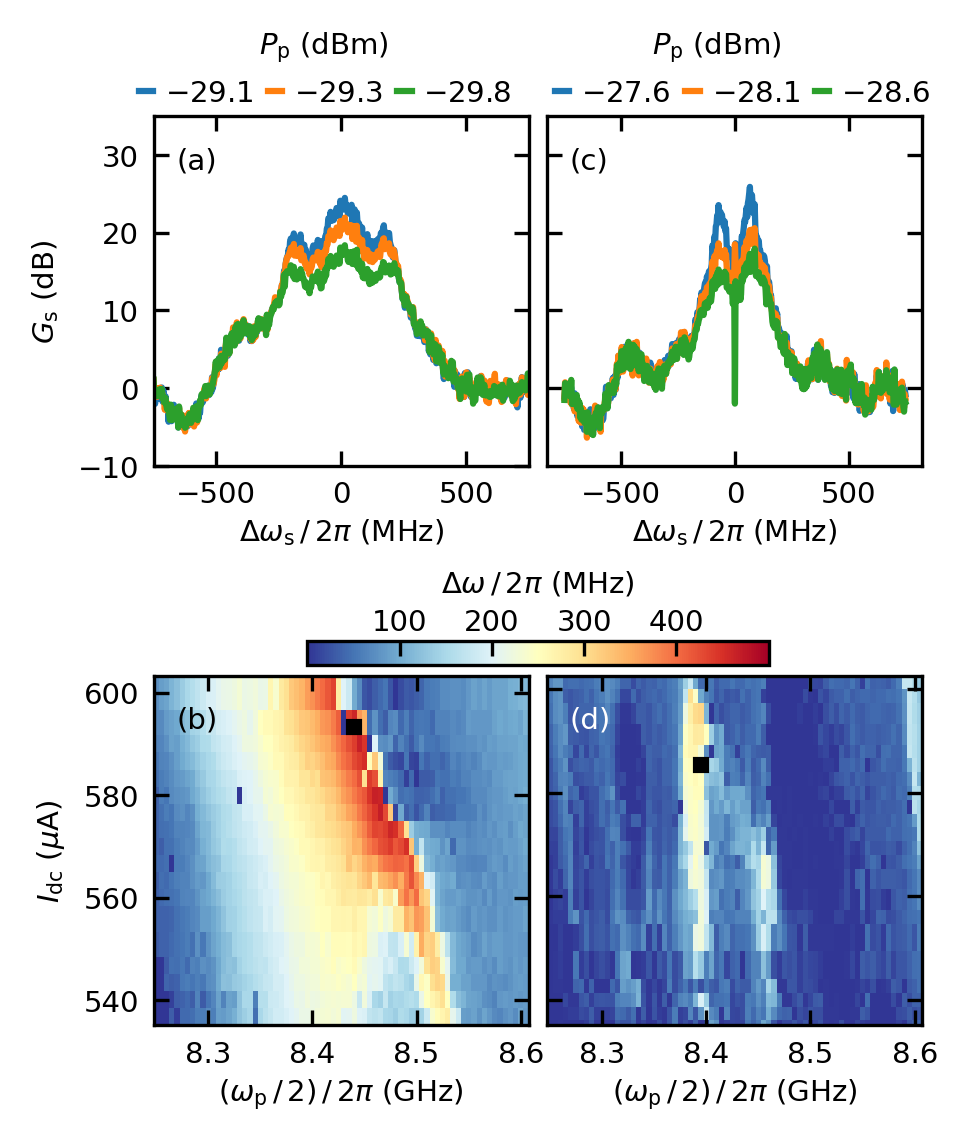}
    \caption{Gain performance with the same measurement procedure as in Fig.~\ref{fig2}(c) and~(e). The data in~(a) and~(b) are measured with the device in Fig.~\ref{fig1} but in a different cooldown. The results in~(c) and~(d) are obtained in a different device.}\label{sfig_diff}
\end{figure}

In different cooldowns, we observe only a-few-MHz variations in the resonance frequency of the nonlinear resonator. 
While the gain profiles (e.g., ripples and fine structures) at a fixed pump frequency and bias current may differ slightly between cooldowns, as shown in Figs.~\ref{sfig_diff}(a) and~(b), the dependence of the bandwidth on the pump frequency and the bias current, as well as the maximum achievable bandwidth, are consistent with the results reported in Fig.~\ref{fig2}.
Although we did not systematically investigate how the superconductivity-breakdown threshold changes across different cooldown cycles, our observations consistently revealed similar critical currents, indicating that the breakdown threshold is relatively constant across cycles.

We have also tested different chips. 
The gain performance of another device with slightly different $Z_{\mathrm{KI},\lambda/4}$ or $Z_{\lambda/4}$ values is shown in Figs.~\ref{sfig_diff}(c) and~(d).
We observed that this device had a less bandwidth (approximately~300 MHz) and a striped characteristic, consistent with a strong impedance mismatch in the environment.

\bibliography{reference}

@article{aumentado2020superconducting,
  title = {{Superconducting parametric amplifiers: The state of the art in Josephson parametric amplifiers}},
  author={Aumentado, Jose},
  journal={IEEE Microwave Magazine},
  volume={21},
  number={8},
  pages={45--59},
  year={2020},
  publisher={IEEE}
}

@article{kaufman2023josephson,
  title={{Josephson parametric amplifier with Chebyshev gain profile and high saturation}},
  author={Kaufman, Ryan and White, Theodore and Dykman, Mark I and Iorio, Andrea and Sterling, George and Hong, Sabrina and Opremcak, Alex and Bengtsson, Andreas and Faoro, Lara and Bardin, Joseph C and others},
  journal={Physical Review Applied},
  volume={20},
  number={5},
  pages={054058},
  year={2023},
  publisher={APS}
}

@article{qing2024broadband,
  title={Broadband coplanar-waveguide-based impedance-transformed Josephson parametric amplifier},
  author={Qing, Bingcheng and Nguyen, Long B and Liu, Xinyu and Ren, Hengjiang and Livingston, William P and Goss, Noah and Hajr, Ahmed and Chistolini, Trevor and Pedramrazi, Zahra and Santiago, David I and others},
  journal={Physical Review Research},
  volume={6},
  number={1},
  pages={L012035},
  year={2024},
  publisher={APS}
}

@article{sarkar2022quantum,
  title={{Quantum-noise-limited microwave amplification using a graphene Josephson junction}},
  author={Sarkar, Joydip and Salunkhe, Kishor V and Mandal, Supriya and Ghatak, Subhamoy and Marchawala, Alisha H and Das, Ipsita and Watanabe, Kenji and Taniguchi, Takashi and Vijay, R and Deshmukh, Mandar M},
  journal={Nature Nanotechnology},
  volume={17},
  number={11},
  pages={1147--1152},
  year={2022},
  publisher={Nature Publishing Group UK London}
}

@article{zorin2016josephson,
  title={{Josephson traveling-wave parametric amplifier with three-wave mixing}},
  author={Zorin, AB},
  journal={Physical Review Applied},
  volume={6},
  number={3},
  pages={034006},
  year={2016},
  publisher={APS}
}

@article{malnou2022performance,
  title={{Performance of a kinetic inductance traveling-wave parametric amplifier at 4 Kelvin: Toward an alternative to semiconductor amplifiers}},
  author={Malnou, Maxime and Aumentado, Joe and Vissers, MR and Wheeler, JD and Hubmayr, Johannes and Ullom, JN and Gao, Jiansong},
  journal={Physical Review Applied},
  volume={17},
  number={4},
  pages={044009},
  year={2022},
  publisher={APS}
}

@article{astafiev2010resonance,
  title={Resonance fluorescence of a single artificial atom},
  author={Astafiev, O and Zagoskin, Alexandre M and Abdumalikov Jr, AA and Pashkin, Yu A and Yamamoto, T and Inomata, K and Nakamura, Y and Tsai, Jaw Shen},
  journal={Science},
  volume={327},
  number={5967},
  pages={840--843},
  year={2010},
  publisher={American Association for the Advancement of Science}
}

@article{callen1951irreversibility,
  title={{Irreversibility and generalized noise}},
  author={Callen, Herbert B and Welton, Theodore A},
  journal={Physical Review},
  volume={83},
  number={1},
  pages={34},
  year={1951},
  publisher={APS}
}

@inproceedings{kerr1997receiver,
  title={{Receiver noise temperature, the quantum noise limit, and the role of the zero-point fluctuations}},
  author={Kerr, Anthony R and Feldman, Marc J and Pan, Shing-Kuo},
  booktitle={Proc. of the 8th Int. Symp. on Space Terahertz Technology},
  pages={101--111},
  year={1997}
}

@article{sivak2020josephson,
  title={{Josephson array-mode parametric amplifier}},
  author={Sivak, VV and Shankar, Shyam and Liu, Gangqiang and Aumentado, Jose and Devoret, MH},
  journal={Physical Review Applied},
  volume={13},
  number={2},
  pages={024014},
  year={2020},
  publisher={APS}
}

@article{yurke1989observation,
  title = {{Observation of parametric amplification and deamplification in a Josephson parametric amplifier}},
  author={Yurke, Bernard and Corruccini, LR and Kaminsky, PG and Rupp, LW and Smith, AD and Silver, AH and Simon, RW and Whittaker, EA},
  journal={Physical Review A},
  volume={39},
  number={5},
  pages={2519},
  year={1989},
  publisher={APS}
}

@article{shu2021nonlinearity,
  title = {{Nonlinearity and wide-band parametric amplification in a (Nb, Ti) N microstrip transmission line}},
  author={Shu, Shibo and Klimovich, Nikita and Eom, Byeong Ho and Beyer, AD and Thakur, R Basu and Leduc, HG and Day, PK},
  journal={Physical Review Research},
  volume={3},
  number={2},
  pages={023184},
  year={2021},
  publisher={APS}
}

@article{mirhosseini2019cavity,
  title = {{Cavity quantum electrodynamics with atom-like mirrors}},
  author={Mirhosseini, Mohammad and Kim, Eunjong and Zhang, Xueyue and Sipahigil, Alp and Dieterle, Paul B and Keller, Andrew J and Asenjo-Garcia, Ana and Chang, Darrick E and Painter, Oskar},
  journal={Nature},
  volume={569},
  number={7758},
  pages={692--697},
  year={2019},
  publisher={Nature Publishing Group UK London}
}

@article{vissers2015frequency,
  title = {{Frequency-tunable superconducting resonators via nonlinear kinetic inductance}},
  author={Vissers, Michael R and Hubmayr, Johannes and Sandberg, Martin and Chaudhuri, Saptarshi and Bockstiegel, Clint and Gao, Jiansong},
  journal={Applied Physics Letters},
  volume={107},
  number={6},
  year={2015},
  publisher={AIP Publishing},
    pages = {062601},

}

@article{naaman2022synthesis,
  title = {{Synthesis of parametrically coupled networks}},
  author={Naaman, Ofer and Aumentado, Jos{\'e}},
  journal={PRX Quantum},
  volume={3},
  number={2},
  pages={020201},
  year={2022},
  publisher={APS}
}

@article{faramarzi20244,
  title = {{A 4-8 GHz Kinetic Inductance Travelling-Wave Parametric Amplifier Using Four-Wave Mixing with Near Quantum-Limit Noise Performance}},
  author={Faramarzi, Farzad and Stephenson, Ryan and Sypkens, Sasha and Eom, Byeong H and LeDuc, Henry and Day, Peter},
  journal={arXiv preprint arXiv:2402.11751},
  year={2024}
}

@article{peng2022floquet,
  title = {{Floquet-mode traveling-wave parametric amplifiers}},
  author={Peng, Kaidong and Naghiloo, Mahdi and Wang, Jennifer and Cunningham, Gregory D and Ye, Yufeng and O’Brien, Kevin P},
  journal={PRX Quantum},
  volume={3},
  number={2},
  pages={020306},
  year={2022},
  publisher={APS}
}

@article{grebel2021flux,
  title={{Flux-pumped impedance-engineered broadband Josephson parametric amplifier}},
  author={Grebel, J and Bienfait, A and Dumur, {\'E} and Chang, H-S and Chou, M-H and Conner, CR and Peairs, GA and Povey, RG and Zhong, YP and Cleland, AN},
  journal={Applied Physics Letters},
  volume={118},
  number={14},
  year={2021},
  publisher={AIP Publishing},
    pages = {142601},

}

@article{ho2012wideband,
  title = {{A wideband, low-noise superconducting amplifier with high dynamic range}},
  author={Ho Eom, Byeong and Day, Peter K and LeDuc, Henry G and Zmuidzinas, Jonas},
  journal={Nature Physics},
  volume={8},
  number={8},
  pages={623--627},
  year={2012},
  publisher={Nature Publishing Group UK London}
}

@article{bockstiegel2014development,
  title = {{Development of a broadband NbTiN traveling wave parametric amplifier for MKID readout}},
  author={Bockstiegel, Clinton and Gao, J and Vissers, MR and Sandberg, M and Chaudhuri, S and Sanders, A and Vale, LR and Irwin, KD and Pappas, DP},
  journal={Journal of Low Temperature Physics},
  volume={176},
  pages={476--482},
  year={2014},
  publisher={Springer}
}

@article{boutin2017effect,
  title = {{Effect of higher-order nonlinearities on amplification and squeezing in Josephson parametric amplifiers}},
  author={Boutin, Samuel and Toyli, David M and Venkatramani, Aditya V and Eddins, Andrew W and Siddiqi, Irfan and Blais, Alexandre},
  journal={Physical Review Applied},
  volume={8},
  number={5},
  pages={054030},
  year={2017},
  publisher={APS}
}

@article{castellanos2007widely,
  title = {{Widely tunable parametric amplifier based on a superconducting quantum interference device array resonator}},
  author={Castellanos-Beltran, MA and Lehnert, KW},
  journal={Applied Physics Letters},
  volume={91},
  number={8},
  year={2007},
  publisher={AIP Publishing},
    pages = {083509},

}

@article{yamamoto2008flux,
  title = {{Flux-driven Josephson parametric amplifier}},
  author={Yamamoto, Tsuyoshi and Inomata, K and Watanabe, M and Matsuba, K and Miyazaki, T and Oliver, William D and Nakamura, Yasunobu and Tsai, JS},
  journal={Applied Physics Letters},
  volume={93},
  number={4},
  year={2008},
  pages = {042510},
  month = {07},
    issn = {0003-6951},
    doi = {10.1063/1.2964182},
    url = {https://doi.org/10.1063/1.2964182},
  publisher={AIP Publishing}
}

@article{ezenkova2022broadband,
  title = {{Broadband SNAIL parametric amplifier with microstrip impedance transformer}},
  author={Ezenkova, D and Moskalev, D and Smirnov, N and Ivanov, A and Matanin, A and Polozov, V and Echeistov, V and Malevannaya, E and Samoylov, A and Zikiy, E and others},
  journal={Applied Physics Letters},
  volume={121 (23)},
  number={23},
  year={2022},
  publisher={AIP Publishing},
    pages = {232601},

}

@article{getsinger1963prototypes,
  title = {{Prototypes for use in broadbanding reflection amplifiers}},
  author={Getsinger, WJ},
  journal={IEEE Transactions on Microwave Theory and Techniques},
  volume={11},
  number={6},
  pages={486--497},
  year={1963},
  publisher={IEEE}
}

@article{sundqvist2014negative,
  title = {{Negative-resistance models for parametrically flux-pumped superconducting quantum interference devices}},
  author={Sundqvist, Kyle M and Delsing, Per},
  journal={EPJ Quantum Technology},
  volume={1},
  pages={1--21},
  year={2014},
  publisher={Springer}
}

@article{macklin2015near,
  title = {{A near-quantum-limited Josephson traveling-wave parametric amplifier}},
  author={Macklin, Chris and O’brien, K and Hover, D and Schwartz, ME and Bolkhovsky, V and Zhang, X and Oliver, WD and Siddiqi, I},
  journal={Science},
  volume={350},
  number={6258},
  pages={307--310},
  year={2015},
  publisher={American Association for the Advancement of Science}
}

@article{clerk2010introduction,
  title = {{Introduction to quantum noise, measurement, and amplification}},
  author={Clerk, Aashish A and Devoret, Michel H and Girvin, Steven M and Marquardt, Florian and Schoelkopf, Robert J},
  journal={Reviews of Modern Physics},
  volume={82},
  number={2},
  pages={1155--1208},
  year={2010},
  publisher={APS}
}

@article{vijay2009invited,
  title = {{Invited Review article: The Josephson bifurcation amplifier}},
  author={Vijay, R and Devoret, MH and Siddiqi, I},
  journal={Review of Scientific Instruments},
  volume={80},
  number={11},
  year={2009},
  publisher={AIP Publishing}
}

@article{niepce2019high,
  title = {{High kinetic inductance NbN nanowire superinductors}},
  author={Niepce, David and Burnett, Jonathan and Bylander, Jonas},
  journal={Physical Review Applied},
  volume={11},
  number={4},
  pages={044014},
  year={2019},
  publisher={APS}
}

@article{makise2010characterization,
  title = {{Characterization of NbTiN thin films deposited on various substrates}},
  author={Makise, Kazumasa and Terai, Hirotaka and Takeda, Masanori and Uzawa, Yoshinori and Wang, Zhen},
  journal={IEEE Transactions on Applied Superconductivity},
  volume={21},
  number={3},
  pages={139--142},
  year={2010},
  publisher={IEEE}
}

@article{frasca2024three,
  title={{Three-wave-mixing quantum-limited kinetic inductance parametric amplifier operating at 6 T near 1 K}},
  author={Frasca, Simone and Roy, Camille and Beaulieu, Guillaume and Scarlino, Pasquale},
  journal={Physical Review Applied},
  volume={21},
  number={2},
  pages={024011},
  year={2024},
  publisher={APS}
}

@article{samkharadze2016high,
  title = {{High-kinetic-inductance superconducting nanowire resonators for circuit QED in a magnetic field}},
  author={Samkharadze, Nodar and Bruno, A and Scarlino, Pasquale and Zheng, G and DiVincenzo, DP and DiCarlo, L and Vandersypen, LMK},
  journal={Physical Review Applied},
  volume={5},
  number={4},
  pages={044004},
  year={2016},
  publisher={APS}
}

@article{vine2023situ,
  title = {{In situ amplification of spin echoes within a kinetic inductance parametric amplifier}},
  author={Vine, Wyatt and Savytskyi, Mykhailo and Vaartjes, Arjen and Kringh{\o}j, Anders and Parker, Daniel and Slack-Smith, James and Schenkel, Thomas and M{\o}lmer, Klaus and McCallum, Jeffrey C and Johnson, Brett C and others},
  journal={Science Advances},
  volume={9},
  number={10},
  pages={eadg1593},
  year={2023},
  publisher={American Association for the Advancement of Science}
}

@article{eichler2017electron,
  title = {{Electron spin resonance at the level of $10^4$ spins using low impedance superconducting resonators}},
  author={Eichler, C and Sigillito, AJ and Lyon, Stephen A and Petta, Jason R},
  journal={Physical Review Letters},
  volume={118},
  number={3},
  pages={037701},
  year={2017},
  publisher={APS}
}

@article{bienfait2016reaching,
  title = {{Reaching the quantum limit of sensitivity in electron spin resonance}},
  author={Bienfait, A and Pla, JJ and Kubo, Y and Stern, M and Zhou, X and Lo, CC and Weis, CD and Schenkel, T and Thewalt, MLW and Vion, D and others},
  journal={Nature Nanotechnology},
  volume={11},
  number={3},
  pages={253--257},
  year={2016},
  publisher={Nature Publishing Group UK London}
}

@article{heinsoo2018rapid,
  title = {{Rapid high-fidelity multiplexed readout of superconducting qubits}},
  author={Heinsoo, Johannes and Andersen, Christian Kraglund and Remm, Ants and Krinner, Sebastian and Walter, Theodore and Salath{\'e}, Yves and Gasparinetti, Simone and Besse, Jean-Claude and Poto{\v{c}}nik, Anton and Wallraff, Andreas and others},
  journal={Physical Review Applied},
  volume={10},
  number={3},
  pages={034040},
  year={2018},
  publisher={APS}
}

@article{zhao2022realization,
  title = {{Realization of an error-correcting surface code with superconducting qubits}},
  author={Zhao, Youwei and Ye, Yangsen and Huang, He-Liang and Zhang, Yiming and Wu, Dachao and Guan, Huijie and Zhu, Qingling and Wei, Zuolin and He, Tan and Cao, Sirui and others},
  journal={Physical Review Letters},
  volume={129},
  number={3},
  pages={030501},
  year={2022},
  publisher={APS}
}

@article{corcoles2015demonstration,
  title = {{Demonstration of a quantum error detection code using a square lattice of four superconducting qubits}},
  author={C{\'o}rcoles, Antonio D and Magesan, Easwar and Srinivasan, Srikanth J and Cross, Andrew W and Steffen, Matthias and Gambetta, Jay M and Chow, Jerry M},
  journal={Nature {Communications}},
  volume={6},
  number={1},
  pages={6979},
  year={2015},
  publisher={Nature Publishing Group UK London}
}

@article{dark_matter_1,
  title = {{A quantum enhanced search for dark matter axions}},
  author={Backes, Kelly M and Palken, Daniel A and Kenany, S Al and Brubaker, Benjamin M and Cahn, SB and Droster, A and Hilton, Gene C and Ghosh, Sumita and Jackson, H and Lamoreaux, Steve K and others},
  journal={Nature},
  volume={590},
  number={7845},
  pages={238--242},
  year={2021},
  publisher={Nature Publishing Group UK London}
}

@article{dark_matter_2,
  title = {{Extended search for the invisible axion with the axion dark matter experiment}},
  author={Braine, Thomas and Cervantes, R and Crisosto, N and Du, N and Kimes, S and Rosenberg, LJ and Rybka, G and Yang, J and Bowring, D and Chou, AS and others},
  journal={Physical Review Letters},
  volume={124},
  number={10},
  pages={101303},
  year={2020},
  publisher={APS}
}

@article{kurpiers2017characterizing,
  title = {{Characterizing the attenuation of coaxial and rectangular microwave-frequency waveguides at cryogenic temperatures}},
  author={Kurpiers, Philipp and Walter, Theodore and Magnard, Paul and Salathe, Yves and Wallraff, Andreas},
  journal={EPJ Quantum Technology},
  volume={4},
  number={1},
  pages={1--15},
  year={2017},
  publisher={Springer}
}

@article{parker2022degenerate,
  title = {{Degenerate parametric amplification via three-wave mixing using kinetic inductance}},
  author={Parker, Daniel J and Savytskyi, Mykhailo and Vine, Wyatt and Laucht, Arne and Duty, Timothy and Morello, Andrea and Grimsmo, Arne L and Pla, Jarryd J},
  journal={Physical Review Applied},
  volume={17},
  number={3},
  pages={034064},
  year={2022},
  publisher={APS}
}

@article{qiu2023broadband,
  title = {{Broadband squeezed microwaves and amplification with a Josephson travelling-wave parametric amplifier}},
  author={Qiu, Jack Y and Grimsmo, Arne and Peng, Kaidong and Kannan, Bharath and Lienhard, Benjamin and Sung, Youngkyu and Krantz, Philip and Bolkhovsky, Vladimir and Calusine, Greg and Kim, David and others},
  journal={Nature Physics},
  volume={19},
  number={5},
  pages={706--713},
  year={2023},
  publisher={Nature Publishing Group UK London}
}

@article{clem2012kinetic,
  title = {{Kinetic impedance and depairing in thin and narrow superconducting films}},
  author={Clem, John R and Kogan, VG},
  journal={Physical Review B},
  volume={86},
  number={17},
  pages={174521},
  year={2012},
  publisher={APS}
}

@article{white2023readout,
  title={Readout of a quantum processor with high dynamic range Josephson parametric amplifiers},
  author={White, Theodore and Opremcak, Alex and Sterling, George and Korotkov, Alexander and Sank, Daniel and Acharya, Rajeev and Ansmann, Markus and Arute, Frank and Arya, Kunal and Bardin, Joseph C and others},
  journal={Applied Physics Letters},
  volume={122},
  number={1},
  year={2023},
  publisher={AIP Publishing}
}

@article{mutus2014strong,
  title = {{Strong environmental coupling in a Josephson parametric amplifier}},
  author={Mutus, Josh Y and White, Theodore C and Barends, Rami and Chen, Yu and Chen, Zijun and Chiaro, Ben and Dunsworth, Andrew and Jeffrey, Evan and Kelly, Julian and Megrant, Anthony and others},
  journal={Applied Physics Letters},
  volume={104},
  number={26},
  year={2014},
  publisher={AIP Publishing},
pages = {263513},

}

@article{PhysRevD.26.1817,
  title = {{Quantum limits on noise in linear amplifiers}},
  author = {Caves, Carlton M.},
  journal = {Phys. Rev. D},
  volume = {26},
  issue = {8},
  pages = {1817--1839},
  numpages = {0},
  year = {1982},
  month = {Oct},
  publisher = {American Physical Society},
  doi = {10.1103/PhysRevD.26.1817},
  url = {https://link.aps.org/doi/10.1103/PhysRevD.26.1817}
}

@article{eichler2011experimental,
  title={{Experimental state tomography of itinerant single microwave photons}},
  author={Eichler, Christopher and Bozyigit, Deniz and Lang, Christian and Steffen, Lars and Fink, Johannes and Wallraff, Andreas},
  journal={Physical Review Letters},
  volume={106},
  number={22},
  pages={220503},
  year={2011},
  publisher={APS}
}

@article{ahn2024extensive,
  title={{Extensive search for axion dark matter over 1 GHz with CAPP’s Main Axion eXperiment}},
  author={Ahn, Saebyeok and Kim, JinMyeong and Ivanov, Boris I and Kwon, Ohjoon and Byun, HeeSu and Van Loo, Arjan F and Park, SeongTae and Jeong, Junu and Lee, Soohyung and Kim, Jinsu and others},
  journal={Physical Review X},
  volume={14},
  number={3},
  pages={031023},
  year={2024},
  publisher={APS}
}

@article{mallet2011quantum,
  title={{Quantum state tomography of an itinerant squeezed microwave field}},
  author={Mallet, F and Castellanos-Beltran, MA and Ku, HS and Glancy, S and Knill, E and Irwin, KD and Hilton, GC and Vale, LR and Lehnert, KW},
  journal={Physical Review Letters},
  volume={106},
  number={22},
  pages={220502},
  year={2011},
  publisher={APS}
}

@article{spring2025fast,
  title={{Fast Multiplexed Superconducting-Qubit Readout with Intrinsic Purcell Filtering Using a Multiconductor Transmission Line}},
  author={Spring, Peter A and Milanovic, Luka and Sunada, Yoshiki and Wang, Shiyu and Van Loo, Arjan F and Tamate, Shuhei and Nakamura, Yasunobu},
  journal={PRX Quantum},
  volume={6},
  number={2},
  pages={020345},
  year={2025},
  publisher={APS}
}

@inproceedings{ranzani2022wideband,
  title = {{Wideband Josephson parametric amplifier with integrated transmission line transformer}},
  author={Ranzani, Leonardo and Ribeill, Guilhem and Hassick, Brian and Fong, Kin Chung},
  booktitle = {{2022 IEEE International Conference on Quantum Computing and Engineering (QCE)}},
  pages={314--319},
  year={2022},
  organization={IEEE}
}

@article{roy2015broadband,
  title = {{Broadband parametric amplification with impedance engineering: Beyond the gain-bandwidth product}},
  author={Roy, Tanay and Kundu, Suman and Chand, Madhavi and Vadiraj, AM and Ranadive, A and Nehra, N and Patankar, Meghan P and Aumentado, J and Clerk, AA and Vijay, R},
  journal={Applied Physics Letters},
  volume={107},
  number={26},
  year={2015},
  publisher={AIP Publishing},
    pages = {262601},

}

@article{devoret1995quantum,
  title = {{Quantum fluctuations in electrical circuits}},
  author={Devoret, Michel H and others},
  journal={Proceedings of the Les Houches
Summer School, Session LXIII, edited by S. Reymaud, E. Giacobino, and J. Zinn-Justin (Elsevier Science, B.V., 1997)},
  volume={7},
  number={8},
  pages={133--135},
  year={1995},
  publisher={Elsevier New York}
}

@article{butseraen2022gate,
  title={{A gate-tunable graphene Josephson parametric amplifier}},
  author={Butseraen, Guilliam and Ranadive, Arpit and Aparicio, Nicolas and Rafsanjani Amin, Kazi and Juyal, Abhishek and Esposito, Martina and Watanabe, Kenji and Taniguchi, Takashi and Roch, Nicolas and Lefloch, Fran{\c{c}}ois and others},
  journal={Nature Nanotechnology},
  volume={17},
  number={11},
  pages={1153--1158},
  year={2022},
  publisher={Nature Publishing Group UK London}
}

@article{frattini2018optimizing,
  title={{Optimizing the nonlinearity and dissipation of a snail parametric amplifier for dynamic range}},
  author={Frattini, NE and Sivak, VV and Lingenfelter, A and Shankar, S and Devoret, MH},
  journal={Physical Review Applied},
  volume={10},
  number={5},
  pages={054020},
  year={2018},
  publisher={APS}
}

@article{remm2023intermodulation,
  title={{Intermodulation distortion in a Josephson traveling-wave parametric amplifier}},
  author={Remm, Ants and Krinner, Sebastian and Lacroix, Nathan and Hellings, Christoph and Swiadek, Fran{\c{c}}ois and Norris, Graham J and Eichler, Christopher and Wallraff, Andreas},
  journal={Physical Review Applied},
  volume={20},
  number={3},
  pages={034027},
  year={2023},
  publisher={APS}
}

@article{dai2024optimizing,
  title={{Optimizing the pump coupling for a three-wave mixing Josephson parametric amplifier}},
  author={Dai, Wei and Liu, Gangqiang and Joshi, Vidul and Miano, Alessandro and Sivak, Volodymyr and Shankar, Shyam and Devoret, Michel H},
  journal={arXiv preprint arXiv:2411.07208},
  year={2024}
}

@article{zmuidzinas2012superconducting,
  title = {{Superconducting microresonators: physics and applications}},
  author={Zmuidzinas, Jonas},
  journal={Annu. Rev. Condens. Matter Phys.},
  volume={3},
  number={1},
  pages={169--214},
  year={2012},
  publisher={Annual Reviews}
}

@book{mazin2005microwave,
  title = {{Microwave kinetic inductance detectors}},
  author={Mazin, Benjamin A},
  year={2005},
  publisher={California Institute of Technology}
}

@article{pappas2011two,
  title = {{Two level system loss in superconducting microwave resonators}},
  author={Pappas, David P and Vissers, Michael R and Wisbey, David S and Kline, Jeffrey S and Gao, Jiansong},
  journal={IEEE Transactions on Applied Superconductivity},
  volume={21},
  number={3},
  pages={871--874},
  year={2011},
  publisher={IEEE}
}

@book{pozar2021microwave,
  title = {{Microwave Engineering: Theory and Techniques}},
  author={Pozar, David M},
  year={2021},
  publisher={John Wiley \& Sons}
}

@inproceedings{mazin2009microwave,
  title={{Microwave kinetic inductance detectors: The first decade}},
  author={Mazin, Benjamin A},
  booktitle={AIP Conference Proceedings},
  volume={1185},
  number={1},
  pages={135--142},
  year={2009},
  organization={American Institute of Physics}
}

@article{vaartjes2024strong,
  title={{Strong microwave squeezing above 1 Tesla and 1 Kelvin}},
  author={Vaartjes, Arjen and Kringh{\o}j, Anders and Vine, Wyatt and Day, Tom and Morello, Andrea and Pla, Jarryd J},
  journal={Nature Communications},
  volume={15},
  number={1},
  pages={4229},
  year={2024},
  publisher={Nature Publishing Group UK London}
}

@article{xu2023magnetic,
  title={{Magnetic field-resilient quantum-limited parametric amplifier}},
  author={Xu, Mingrui and Cheng, Risheng and Wu, Yufeng and Liu, Gangqiang and Tang, Hong X},
  journal={PRX Quantum},
  volume={4},
  number={1},
  pages={010322},
  year={2023},
  publisher={APS}
}

@article{khalifa2023nonlinearity,
  title={{Nonlinearity and parametric amplification of superconducting nanowire resonators in magnetic field}},
  author={Khalifa, Mohammad and Salfi, Joseph},
  journal={Physical Review Applied},
  volume={19},
  number={3},
  pages={034024},
  year={2023},
  publisher={APS}
}

@article{khalifa2024kinetic,
  title={{Kinetic inductance parametric converter}},
  author={Khalifa, M and Feldmann, P and Salfi, J},
  journal={Physical Review Applied},
  volume={22},
  number={2},
  pages={024025},
  year={2024},
  publisher={APS}
}

@article{splitthoff2024gate,
  title={{Gate-tunable kinetic inductance parametric amplifier}},
  author={Splitthoff, Lukas Johannes and Wesdorp, Jaap Joachim and Pita-Vidal, Marta and Bargerbos, Arno and Liu, Yu and Andersen, Christian Kraglund},
  journal={Physical Review Applied},
  volume={21},
  number={1},
  pages={014052},
  year={2024},
  publisher={APS}
}

@article{zapata2024granular,
  title={{Granular aluminum parametric amplifier for low-noise measurements in tesla fields}},
  author={Zapata, Nicolas and Takmakov, Ivan and G{\"u}nzler, Simon and Geisert, Simon and Ihssen, Soeren and Field, Mitchell and Nambisan, Ameya and Rieger, Dennis and Reisinger, Thomas and Wernsdorfer, Wolfgang and others},
  journal={Physical Review Letters},
  volume={133},
  number={26},
  pages={260604},
  year={2024},
  publisher={APS}
}

@article{mohamed2024selective,
  title={{Selective single-and double-mode quantum-limited amplifier}},
  author={Mohamed, Abdul and Zohari, Elham and Pla, Jarryd J and Barclay, Paul E and Barzanjeh, Shabir},
  journal={Physical Review Applied},
  volume={21},
  number={6},
  pages={064052},
  year={2024},
  publisher={APS}
}

@article{lu2022broadband,
  title={{Broadband Josephson parametric amplifier using lumped-element transmission line impedance matching architecture}},
  author={Lu, Yapeng and Xu, Wenqu and Zuo, Quan and Pan, Jiazheng and Wei, Xingyu and Jiang, Junliang and Li, Zishuo and Zhang, Kaixuan and Guo, Tingting and Wang, Shuo and others},
  journal={Applied physics letters},
  volume={120},
  number={8},
  year={2022},
  publisher={AIP Publishing}
}
\end{document}